\documentclass[journal,comsoc]{IEEEtran}

\usepackage[T1]{fontenc}
\usepackage{multirow}

\ifCLASSINFOpdf
\else
\fi

\usepackage{amsmath}

\usepackage[cmintegrals]{newtxmath}

\usepackage{cite}
\usepackage{academicons}
\usepackage{hyperref}
\usepackage{xcolor}
\usepackage{graphicx}
\usepackage{lipsum}
\usepackage{algorithm}
\usepackage{algpseudocode}
\usepackage{tabularx,booktabs}
\usepackage{subfigure}
\usepackage{float}
\usepackage{diagbox}
\usepackage{amsmath}

\usepackage{titlesec}

\begin{document}



\title{Spectral and spatial power evolution design with machine learning-enabled Raman amplification}
        
\author{Mehran~Soltani \href{https://orcid.org/0000-0002-5831-1296}{\includegraphics[scale=0.75]{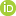}}, Francesco~Da~Ros \href{http://orcid.org/0000-0002-9068-9125}{\includegraphics[scale=0.75]{figures/ORCIDiD_icon16x16.png}}, Andrea~Carena \href{http://orcid.org/0000-0001-6848-3326}{\includegraphics[scale=0.75]{figures/ORCIDiD_icon16x16.png}}, Darko~Zibar \href{http://orcid.org/0000-0003-4182-7488}{\includegraphics[scale=0.75]{figures/ORCIDiD_icon16x16.png}}      
\thanks{M. Soltani, F. Da Ros and D. Zibar are with the Department of Photonics engineering, Technical university of Denmark (DTU Fotonik), Denmark}
\thanks{A. Carena is with the Dipartimento di Elettronica e Telecomunicazioni at Politecnico di Torino, Italy}
}

\markboth{Journal of \LaTeX\ Class Files,~Vol.~14, No.~8, August~2015}%
{Shell \MakeLowercase{\textit{et al.}}: Bare Demo of IEEEtran.cls for IEEE Communications Society Journals}

\maketitle

\begin{abstract}

  We present a machine learning (ML) framework for designing desired signal power profiles over the spectral and spatial domains in the fiber span. The proposed framework adjusts the Raman pump power values to obtain the desired two-dimensional (2D) profiles using a convolutional neural network (CNN) followed by the differential evolution (DE) technique. The CNN learns the mapping between the 2D profiles and their corresponding pump power values using a data-set generated by exciting the amplification setup. Nonetheless, its performance is not accurate for designing 2D profiles of practical interest, such as a 2D flat or a 2D symmetric (with respect to the middle point in distance). To adjust the pump power values more accurately, the DE fine-tunes the power values initialized by the CNN to design the proposed 2D profile with a lower cost value. In the fine-tuning process, the DE employs the direct amplification model which consists of 8 bidirectional propagating pumps, including 2 second-order and 6 first order, in an 80 km fiber span. We evaluate the framework to design broadband 2D flat and symmetric power profiles, as two goals for wavelength division multiplexing (WDM) system performing over the whole C-band. Results indicate the framework's ability to achieve maximum power excursion of 2.81 dB for a 2D flat, and maximum asymmetry of 14\% for a 2D symmetric profile. 

\end{abstract}


\IEEEpeerreviewmaketitle

\section{Introduction}

\IEEEPARstart{D}{\lowercase{istributed}} Raman amplification provides the opportunity to design the signal power evolution jointly in spectral and spatial domains. This amplification scheme offers variety of attractive features such as low noise properties \cite{Agrawal, 7163281}, broad range of gain due to multi-pumping configuration and flexibility in gain and power evolution design \cite{deMoura:21, 8894395}. One of the main challenges with Raman amplifier design is to select the pump powers and their frequency values to achieve desired signal power profiles in the fiber span. Recently, most of the focus has been on designing a desired gain spectra at the receiver side using machine learning and optimization techniques \cite{ 8320966, 8894395, deMoura:20, 9244561 }. In \cite{8320966} a neural network (NN) is combined with genetic algorithm to design flat gain profiles. The authors of \cite{8894395, deMoura:20} propose a NN model to learn the inverse mapping between the gain profiles and the pump powers and wavelengths values. Additionally, an extra NN model is proposed in \cite{8894395, deMoura:20} to fine-tune the pump parameter values through back-propagation. In \cite{9244561}, a differentiable Raman amplification model is presented in the training procedure of NN to predict the pump parameters to target a group of flat or tilted gains.



Besides gain spectra shaping, designing a 2D signal power evolution profile jointly in spectrum and space is a beneficial way to satisfy some of the goals in an optical communication system like enhancing the signal-to-noise ratio (SNR) and mitigation of nonlinear impairments \cite{959353}. For example, distribution of the power evenly in frequency and distance along the fiber resembles a link with effective zero attenuation and results in the minimum accumulated amplified spontaneous emission (ASE) noise \cite{Ania-Castanon:04, 1601058, PhysRevLett.101.123903}. Additionally, a zero-loss medium is a theoretical prerequisite for the transmission based on Nonlinear Fourier Transform (NFT) which relies on analytically solving a lossless nonlinear Schr{\"o}dinger equation (NLSE) \cite{Mollenauer:88, le2015nonlinear}. Alternatively, a possibly less challenging and more realistic target profile in distance for nonlinearity compensation is a symmetric profile. A symmetric power profile with respect to the middle point in distance is required to maximize the performance of an optical phase conjugation (OPC) \cite{Jansen:06} in combating the  nonlinear impairments in the system \cite{Phillips:14, tan2018distributed}.

It has been numerically and experimentally proven that a desired 2D signal power evolution can be achieved using high-order Raman pumps in a bidirectional propagating scheme \cite{Chestnut:02}. Most common Raman amplification setups for designing the power evolution in fiber distance are made by combining second-order and first-order pumps in a bidirectional propagating scheme \cite{Chestnut:02}, or by using Raman pumps combined with fiber bragg gratings (FBGs) as reflectors \cite{Ania-Castanon:04}. In \cite{Ania-Castanon:04, PhysRevLett.101.123903}, an amplification structure based on second-order Raman pumps with FBGs has been presented to provide a quasi-lossless transmission medium by implementing an ultra long Raman fiber laser. Moreover, in \cite{Rosa:15}, numerical optimization of narrow-band signal power asymmetry for OPC has been addressed by considering different amplification setups using first-order and second-order pumps.

Most of the works presented for designing signal power profiles have addressed single-channel narrow-band signals, with a few works approaching a multi-channel quasi-lossless signal transmission \cite{1561354, PhysRevLett.101.123903}. In \cite{1561354, PhysRevLett.101.123903}, the authors present an experimental setup including second-order pumps combined with the FBG reflectors to turn the span into an ultralong laser and achieve a lossless transmission in frequency and distance. The pump power values in this setup need to be tuned heuristically to provide a stable and flat signal power evolution. This process becomes challenging when the target is a desired 2D power profile of practical interest.

\begin{figure*}[!ht]
\centering
\includegraphics[width=\textwidth]{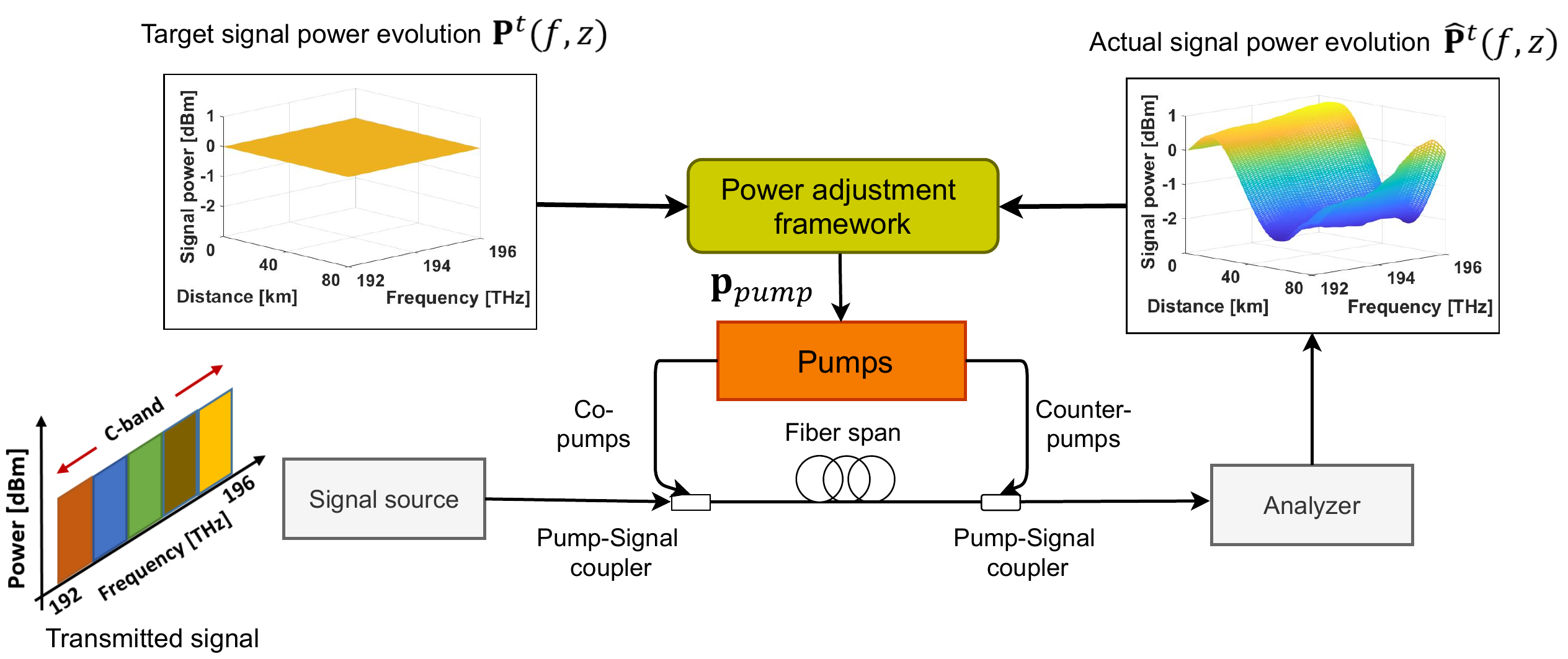}
\caption{Inverse system model applied to adjust the Raman pump powers values in an amplification to design a desired 2D signal power profile.}
\label{fig:general amplification}
\end{figure*}

In this paper we present a ML-based framework to design a desired 2D signal power evolution in wide-band spectrum and along the fiber span by adjusting Raman pump powers values. The proposed framework extends our presented CNN model \cite{Soltani:21} which approximates the mapping between the 2D power profiles and their corresponding Raman pump powers values. The CNN model is trained on a data-set generated by exciting the amplification model with randomly selected pump powers values. In spite of this, its performance is not accurate for designing specific 2D target profiles. In addition, 2D profiles such as a 2D flat or a 2D symmetric have specific and non-differentiable cost functions, which cannot be easily approached with learning-based models like the CNN. To obtain more accurate pump powers values for these profiles, we consider the predicted values by the CNN as an initial solution and apply differential evolution (DE) as a gradient-free fine-tuning technique. The proposed CNN-assisted DE framework uses a direct functional description of the amplification model and fine-tunes the pump power values to design a 2D profile with lower cost value based on the desired objective. The framework is evaluated on designing 2D flat and symmetric power profiles, as two highly addressed targets of practical interest in the literature \cite{Ania-Castanon:04, Rosa:15}.


Alternative to the first-order Raman amplification setup in \cite{Soltani:21}, we propose a higher-order pumping scheme to enable the framework to approach desired 2D profiles. More specifically, we use a bidirectional amplification setup with two second-order and six first order pumps (one second-order and three first-order for each propagation direction) to cover the whole C-band without adding system complexity such as FBGs \cite{Ania-Castanon:04}. Moreover, we assume fixed wavelengths values for the Raman pumps same as in the experiments, where the pump module usually allows to only set the power values. Therefore, the 2D profile design problem will be confined to the prediction of Raman pump powers values.

The remainder of the paper is organized as follows: In section II, we describe the proposed amplification structure followed by the CNN-assisted DE framework for 2D power evolution design problem. In section III, first, the performance of the inverse CNN model is evaluated on designing achievable power profiles. Afterward, the proposed CNN-assisted DE is applied to design 2D flat and symmetric power profiles. Finally, we conclude the paper in section IV.

\section{2D signal power evolution design}
In this section, first, we will describe a general view of a Raman amplification scheme which utilizes a Raman pump power adjustment framework to design a 2D power evolution profile. Afterward, we will present our suggested framework to design 2D spectral-spatial profiles in more details. 

\subsection{Raman pump power adjustment framework}

Fig. \ref{fig:general amplification} illustrates a schematic of a fiber link in which a target 2D signal power profile in spectrum and fiber distance is designed by adjusting the Raman pump power values. In this scheme, a 2D target signal power evolution $\textbf{P}^t(f,z)$ is used as the input to the framework which predicts the Raman pump power values $\textbf{p}_{pump}$, including both co- and counter-propagating pumps. 
The predicted pump powers values are then applied by the Raman pumps into the fiber span where a WDM frequency comb signal with flat spectra in C-band is propagated. 

Due to the stimulated Raman scattering (SRS) phenomenon, the fiber is utilized as the gain medium. Consequently, the signal power evolution in both frequency and distance is tailored by the SRS and a two dimensional matrix referred to as the actual 2D signal power evolution $\hat{\textbf{P}}^t(f,z)$ is calculated. Afterward, $\hat{\textbf{P}}^t(f,z)$ is compared to $\textbf{P}^t(f,z)$ to calculate the corresponding cost value based on the desired objective. The pump power adjustment framework evaluates the cost value and applies an updated set of Raman pump power values to the system aiming to produce a new 2D signal power profile with a lower cost value. This process, which is referred to as "fine-tuning", continues until a certain convergence criteria such as maximum number of iterations or a certain threshold for cost value is satisfied.

As detailed in section I, the fine-tuning models presented in the literature for Raman amplification design have addressed gain spectrum shaping at the receiver \cite{8894395,deMoura:20}. These frameworks employ a NN model to approximate the "direct models" of the amplification system. The direct model is used to evaluate the gain spectrum given the pump parameters values. The cost function in these frameworks is defined as mean square error (MSE) which is differentiable with respect to the model parameters and enables to fine-tune the pump parameters values through back-propagation.

Addressing the target 2D power evolution profiles of practical interest, cost functions for each target may be non-differentiable, and also target-dependent. For instance, for 2D targets that the objective is to minimize the power excursion (also called flatness) or asymmetry in distance \cite{PhysRevLett.101.123903, Quasi-lossless1, Rosa:15}, it is difficult or even may not be possible to calculate a closed-form formula for the differential of the cost function with respect to the model parameters. Additionally, it is shown that mean square error (MSE), despite its simplicity and mathematically convenient form, is not adequate for the assessment of data resembled as 2D images \cite{Wang2004ImageQA}. Considering these issues, we model the pump power adjustment process for designing target 2D practical power profiles as a gradient-free global optimization problem. The proposed framework has the flexibility to design 2D practical power profiles with different corresponding forms of objectives.

\subsection{CNN-assisted DE for 2D power evolution design}

In the proposed framework, we choose DE algorithm, known as an evolutionary-based technique to adjust the pump powers values. DE is a gradient-free technique with the robustness and the flexibility to capture solutions of complex optimization problems \cite{Blum2012}. More importantly, it has been successfully applied to a wide variety of problems due to its strong global search capability, easy implementation and quick convergence \cite{5601760}. 

Generally, evolutionary algorithms like DE start with a population of individuals (as the initial solutions) and try to improve each particle of the population in an iterative process to achieve a point with minimum cost value. In the absence of a prior information about the initial solution, they always start with a random guess in the parameter space \cite{WHITLEY2001817}. Considering this, a requirement of the DE for the proposed 2D power evolution design is to define the search region of pump power values by setting their lower-bound $\textbf{p}_{LB}$ and the upper-bound $\textbf{p}_{UB}$ values. A possible way of specifying $\textbf{p}_{LB}$ and $\textbf{p}_{UB}$ is to set them as the minimum and maximum possible values of pump powers, respectively. This approach is proposed generally when there is no prior information provided about the possible sub-region of the pump power values including the point with minimum cost value. With this blind search approach, the model would be liable to slow convergence, and moreover, plunging into local minimum. This issues are more dominant when the number of pumps, being the dimension of the space to search, increases \cite{app8101945}.


We improve our chance to approach a more reliable and accurate solution for the proposed 2D profile design problem using a ML-based model. The role of the ML model is to provide an initial information regarding a sub-region in the space of pump power values to initialize the DE population. For this purpose, we use our previously presented CNN model in \cite{Soltani:21} to approximate the mapping between the 2D power profiles and their corresponding pump power values. 

The CNN model in \cite{Soltani:21} is a neural network trained based on a data-set generated by random selection of the pump powers values in the proposed amplification setup. It consists of two networks trained end-to-end, a feature extraction and a regression network. The feature extraction network consists of three CNN layers each followed by pooling layers, while the regression network is made up of 2 fully-connected (FC) layers. The role of the first network is to extract the informative and low-dimensional features of the 2D profiles. In the meantime, the second network maps the extracted features to the corresponding pump parameter values. The training process of the CNN aims to update the network weights to minimize the MSE between the true pump powers values and the predicted ones. The CNN has shown a good performance in designing achievable 2D power evolution profiles which are recorded by random selection of the pump powers values in the amplification setup. 

Compared to the achievable set of 2D profiles, the profiles of practical interest are more challenging to obtain due to physical limitations of the amplification systems. Additionally, some practical profiles have strict objectives such as minimum power excursion or minimum asymmetry which cannot be easily approached with the CNN model trained to minimize a loss function such as the MSE. Therefore, in the proposed framework, we employ the CNN to provide a set of pump powers values to be used as an initial solution. This solution gives a prior information about a space region of pump power values where a better solution to the problem falls inside. By initializing the DE population based on the CNN results, we improve the convergence speed and also the quality of the final solution. Considering this, the fine-tuning process is performed with parameter constraints on a relatively tight space region around the set of values predicted by the CNN model.

\begin{figure*}[ht!]
\centering
\includegraphics[width=\textwidth]{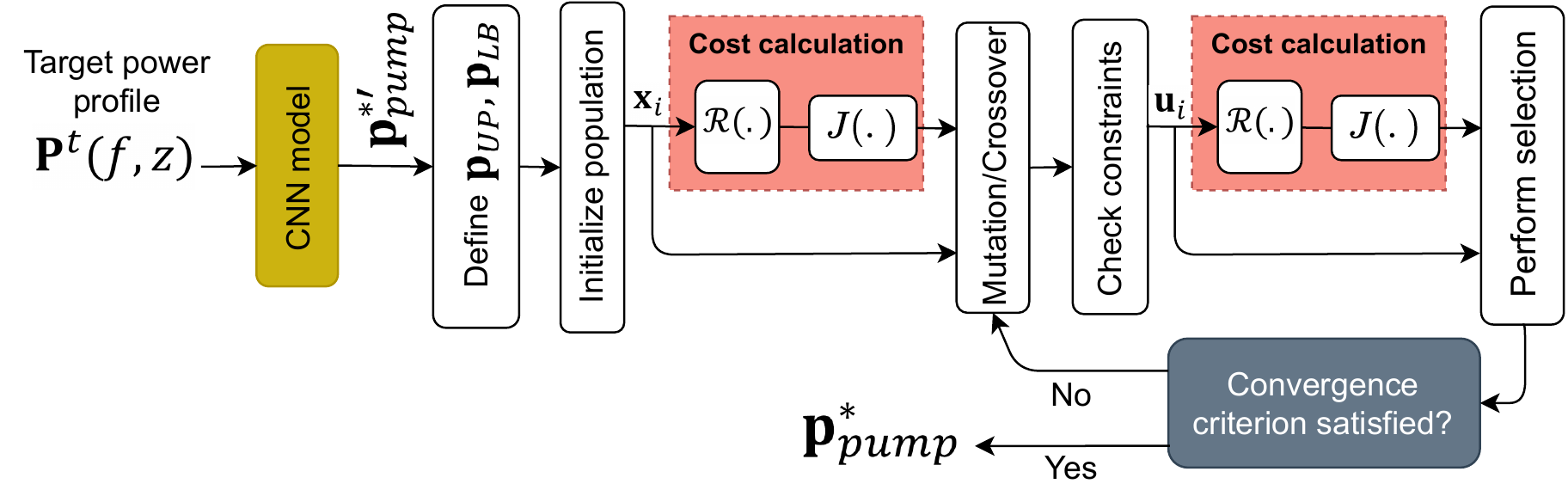}
\caption{Block diagram of the proposed CNN-assisted DE method used for pump power prediction based on a specific 2D target power profile.}
\label{fig:mainmethod}
\end{figure*}

Alternative to the approximate NN-based direct models used for fine-tuning in \cite{8894395, deMoura:20}, we employ the DE combined with the numerically provided direct model of the Raman amplification scheme \cite{Agrawal}. We refer to the functional model described between the pump powers values and their corresponding 2D power profiles as Raman solver $\mathcal{R}(\cdot)$. The 2D power evolution generated by the Raman solver with a given set of pump power values $\textbf{p}_{pump}$ is defined as:

\begin{equation}
\label{eq:Raman}
    \textbf{P}(f,z|\textbf{p}_{pump}) = \mathcal{R}(\textbf{p}_{pump})
\end{equation}

Indeed, the $\mathcal{R}(\cdot)$ function is a set of nonlinear differential equations describing the Raman effect on the signal power evolution \cite{Agrawal}. This set of equations is generally considered as a boundary value problem (BVP) which is solved numerically using ordinary differential equation (ODE) techniques\cite{Ascher}. 


Diagram of the proposed Raman pump power adjustment scheme referred to as "CNN-assisted DE", is shown in Fig. \ref{fig:mainmethod}. First, the 2D target power profile $\textbf{P}^t(f,z)$ is used as the input to the trained CNN model to predict the corresponding Raman pump power values as the initial solution $\textbf{p}_{pump}^{*'} = [p_1, ..., p_{n}]$, where $n$ is the number of pumps. Afterward, the constraints $\textbf{p}_{LB} = \textbf{p}_{pump}^{*'} - \boldsymbol\Delta_p \cdot \textbf{p}_{pump}^{*'}$ and $ \textbf{p}_{UB} = \textbf{p}_{pump}^{*'} + \boldsymbol\Delta_p \cdot \textbf{p}_{pump}^{*'}$ are defined, where $\boldsymbol\Delta_p = [\Delta_1, \Delta_2, ..., \Delta_{n}]$ is a hyper-parameter vector which consists of $n$ number of unit-less scalar values. Each element of $\boldsymbol\Delta_p$ is multiplied by its corresponding initially pump power value in $\textbf{p}_{pump}^{*'}$. The resulting number for each pump power is considered as its amount of deviation and the constraints $\textbf{p}_{LB}$ and $ \textbf{p}_{UB}$, accordingly.


Once the constraints are defined, an initial population with $N_p$ number of individuals is generated from the uniform distribution over the specified ranges. Each individual in the population is represented by a vector $\textbf{x}_{i} = (x_{i}^1, x_{i}^2, ..., x_{i}^{n}), i=1,..., N_p$ as a set of pump power values, which is a candidate for the fine-tuning solution. All of the individuals are passed through the Raman solver in Eq. \ref{eq:Raman} and their corresponding 2D profiles are generated. The resulting 2D profile for each candidate is used further for the calculation of the corresponding pre-defined cost value. The cost value of the $i$th individual $J(\textbf{x}_{i})$ can be defined either as an error between the 2D target $\textbf{P}^t(f,z)$ and the resulting 2D profile $\hat{\textbf{P}}(f,z|{\textbf{x}_i}) = \mathcal{R}(\textbf{x}_{i})$, or based on a metric measured for the resulting 2D profile such as its power excursion or asymmetry value.

For each generated individual $\textbf{x}_{i}$, an iterative DE process including mutation, crossover and evaluation is performed. First, three individuals $\textbf{x}_{r_1}$, $\textbf{x}_{r_2}$ and $\textbf{x}_{r_3}$ are selected randomly from the population where $r_1 \neq r_2 \neq r_3 \neq i$, and a mutation is applied to generate a donor vector $\textbf{v}_{i}$ with the same size as $\textbf{x}_{i}$. The donor vector $\textbf{v}_{i}$ aims to increase the diversity of the population in order to avoid plunging into a local minimum, and it is defined as the following:

\begin{equation}
 \textbf{v}_{i} = \textbf{x}_{r_1} + F\cdot (\textbf{x}_{r_2} - \textbf{x}_{r_3})
\label{donor}
\end{equation}

where $F\in [0,1]$ is the mutation factor, which controls the diversity of the population.

After the mutation, the crossover process develops trial vector $\textbf{u}_{i}$ by combining the elements of the donor vector $\textbf{v}_{i}$ and the current target vector $\textbf{x}_{i}$ as follows:

\begin{equation}
\textbf{u}_{i}^j = 
\begin{cases}
  v_{i}^j, & \text{if}\ r_{i}^j \leq CR \: or \: j = j_{rand} \\
  x_{i}^j, & \text{otherwise}
\end{cases}
\label{crossover}
\end{equation}

\begin{algorithm}[hbt!]
     
    \caption{CNN-assisted DE}\label{euclid}
    \textbf{Input}: Target power profile $\textbf{P}^t(f,z)$, $\boldsymbol\Delta_p = [\Delta_1, \Delta_2, ..., \Delta_{N_p}]$, DE parameters: \{$N_p$: Population size, \textit{CR}: Crossover probability, \textit{F}: Mutation factor, $MaxEv$: Maximum number of cost function evaluations\} \\
    \textbf{Output}: Pump power values $\textbf{p}_{pump}^{*}$\\
    \textbf{Optimization procedure}:
    \begin{algorithmic}[1]
    \State $\textbf{p}_{pump}^{*'} =  CNN(\textbf{P}^t(f,z))$
    \State set $\textbf{p}_{LB} = \textbf{p}_{pump}^{*'} - \boldsymbol\Delta_p \cdot \textbf{p}_{pump}^{*'T}$, $ \textbf{p}_{UB} = \textbf{p}_{pump}^{*'} + \boldsymbol\Delta_p \cdot \textbf{p}_{pump}^{*'T}$
    \State \textbf{Generate population}:  $\textbf{x}_{i} = (x_{i}^1, x_{i}^2, ..., x_{i}^{n}), i=1,..., N_p$, $\textbf{p}_{UB}< \textbf{x}_i <\textbf{p}_{LB}$
    \State {\textbf{Direct model}}: Calculate $\textbf{P}(f,z|\textbf{x}_i)= \mathcal{R}(\textbf{x}_{i})$
    \State \textbf{Calculate individual cost}: $J(\textbf{x}_i)$
    \State \textbf{While} $Ev < MaxEv$:
    \State \hspace*{1.5em} \textbf{for} $i = 1:N_p$  \textbf{do }
    \State \hspace*{3em} {\textbf{Select} three random individuals $x_{r1}, x_{r2}, x_{r3}$ from 
    \hspace*{3em} the population where $i \neq r_1 \neq r_2 \neq r_3$}
    \State \hspace*{3em} {\textbf{Mutation}} : Form the donor vector as in Eq. \ref{donor}
    \State \hspace*{3em} {\textbf{Crossover}}: Form the trial vector as in Eq. \ref{crossover}
    \State \hspace*{3em} {\textbf{Check constraints}}: Check if $\textbf{p}_{LB}< \textbf{u}_{i}<\textbf{p}_{UB}$, 
    \hspace*{4.5em} otherwise, go to the next individual.
    \State \hspace*{3em} {\textbf{Direct model}}: $\textbf{P}(f,z|\textbf{u}_i)= \mathcal{R}(\textbf{u}_{i})$
    \State \hspace*{3em} {\textbf{Evaluate}}: if $J(\textbf{u}_{i}) \leq J(\textbf{x}_{i})$, replace $\textbf{x}_{i}$ with 
    \hspace*{4.5em} $\textbf{u}_{i}$
    \State \hspace*{3em} $Ev \gets Ev + 1$ 
    
    \State \hspace*{1.5em} \textbf{End for}
    \State \textbf{End while}

    \end{algorithmic}
\label{alg:cap}
\end{algorithm}

where $u_{i}^j$, $v_{i}^j$ and $x_{i}^j$ represent the $j$th element of $\textbf{u}_{i}$, $\textbf{v}_{i}$ and $\textbf{x}_{i}$, respectively. $CR$ is the crossover probability,  $r_{i}^j \sim \textit{U}(0,1) $ is a uniform distribution which is generated for each $j$, and $j_{rand} \in {1, 2, ..., N_p}$ is a random integer used to ensure that $\textbf{x}_{i} \neq \textbf{u}_{i}$. After performing the mutation and the crossover, the trial vector is checked if it follows the constraints $\textbf{p}_{LB}$ and $\textbf{p}_{UB}$. Afterward, $\textbf{u}_{i}$ is passed through the Raman solver function to generate the corresponding power evolution profile $\textbf{P}(f,z|\textbf{u}_{i})=\mathcal{R}(\textbf{u}_{i})$ and calculate the cost value $J$, accordingly. The cost values of the trial vector $J(\textbf{u}_{i})$ and the target vector $J(\textbf{x}_{i})$ will be compared and the one with lower cost value will remain in the population. The selection process can be formulated to form the new particle $\textbf{x}_{i_{new}}$ as follows:

\begin{equation}
\textbf{x}_{i_{new}} = 
\begin{cases}
  \textbf{u}_{i}, & \text{if}\ J(\textbf{u}_{i})<J(\textbf{x}_{i}) \\
  \textbf{x}_{i}, & \text{otherwise}
\end{cases}
\end{equation}

The aforementioned process continues further until a convergence criteria is satisfied either by reaching the maximum number of iterations, or by achieving a cost function threshold. In this paper we consider the maximum possible number of cost function evaluations $MaxEv$ as the convergence criteria. The implementation procedure of the proposed CNN-assisted DE optimization scheme is summarized in Algorithm \ref{alg:cap}.


\section{Simulation results}

We test the ability of the proposed framework to provide accurate pump power predictions for 2D signal power evolution profiles. The Raman solver defined in Eq. \ref{eq:Raman} is provided by GNPy library \cite{Gnpy}. This function is utilized for the generation of the data-set for training, testing and validation of the CNN. Moreover, it is used in the cost function calculation process of the CNN-assisted DE framework illustrated in Algorithm \ref{alg:cap}.

We perform the power evolution design procedure jointly in spectrum and spatial domains for a single span of a single mode-fiber (SMF). The proposed pumping scheme is based on 8 pumps, including four counter- and four co- propagating pumps. In order to enable the framework to approach 2D power profiles of practical interest over the whole C-band, each set of co- and counter-propagating pumps consists of one second-order pump with the wavelength of 1366 nm and three first-order pumps with the wavelengths of 1425 nm, 1455 nm and 1475 nm, respectively. Ranges of the pump powers for generation of the data-set are presented in table \ref{pump ranges}.

\begin{table}[h]
\caption{Raman pump wavelengths and power ranges}
\label{pump ranges}
\setlength{\tabcolsep}{2pt} 
\begin{tabular}{m{2.5cm} |m{1.5cm} |m{1.3cm} |m{1.3cm} |m{1.3cm} }
    \hline
    Co-pumps   & $p_1$ & $p_2$ & $p_3$ & $p_4$ \\ \hline
    wavelength [nm]  & 1366 & 1425 & 1455 & 1475 \\
    power range [mW]  & [200-1200]  & [5-150] & [5-150]& [5-150]\\ \hline
    \hline
    Counter-pumps   & $p_5$ & $p_6$ & $p_7$ & $p_8$ \\ \hline
    wavelength [nm] & 1366 & 1425 & 1455 & 1475 \\
    power range [mW] & [200-1200] & [5-150]& [5-150]& [5-150] \\ 

    \hline
\end{tabular}

\end{table}

The 2D signal power profiles are analyzed over the whole C-band (between 192 THz and 196 THz) divided into 40 channels each with a bandwidth of 100 GHz. Input signal power per channel is set to $0$ dBm resulting in the total signal power of $16$ dBm. Moreover, the utilized SMF has the following specifications: length of the span $L_{span} =$ 80 km, signal data attenuation of $\alpha_s = $ 0.2 dB/km, second-order pumps $[p_1, p_5]$ attenuation $\alpha_P^{2nd} =$  0.32 dB/km, and first-order pumps $[p_2, p_3, p_4, p_6, p_7, p_8]$ attenuation $\alpha_P^{1st} = $ 0.25 dB/km, effective area $A_{eff}=$ 80 $\mu m^2$, non-linear coefficient $\gamma=$ 1.26 1/W/km, and Raman coefficient $g_R=$ 0.4125 1/W/km. Standard silica Raman efficiency profile is assumed \cite{Agrawal}. 

The GnPy library \cite{Gnpy} employs a BVP solver based on residual error control to solve the set of Raman differential equations. Considering this, we set the distance resolution to $z_{res} = $ 500 m which satisfies the residual error threshold $r_{error} =10^{-6}$. Since the selected distance step size is lower than the 1 km value reported in \cite{Soltani:21}, the resulting 2D profiles have a larger dimension size compared to the ones in \cite{Soltani:21}. To have extra dimension reduction, we use four CNN layers rather than three layers proposed in the main CNN architecture. The other structural parameters of the network such as the number of the filters and their kernel size, sizes of the pooling layers and the number of nodes for the hidden layers are the same as reported in \cite{Soltani:21}. 

\subsection{Evaluation of the CNN on achievable profiles}
\begin{figure}
\centering
\includegraphics[width=8.5cm]{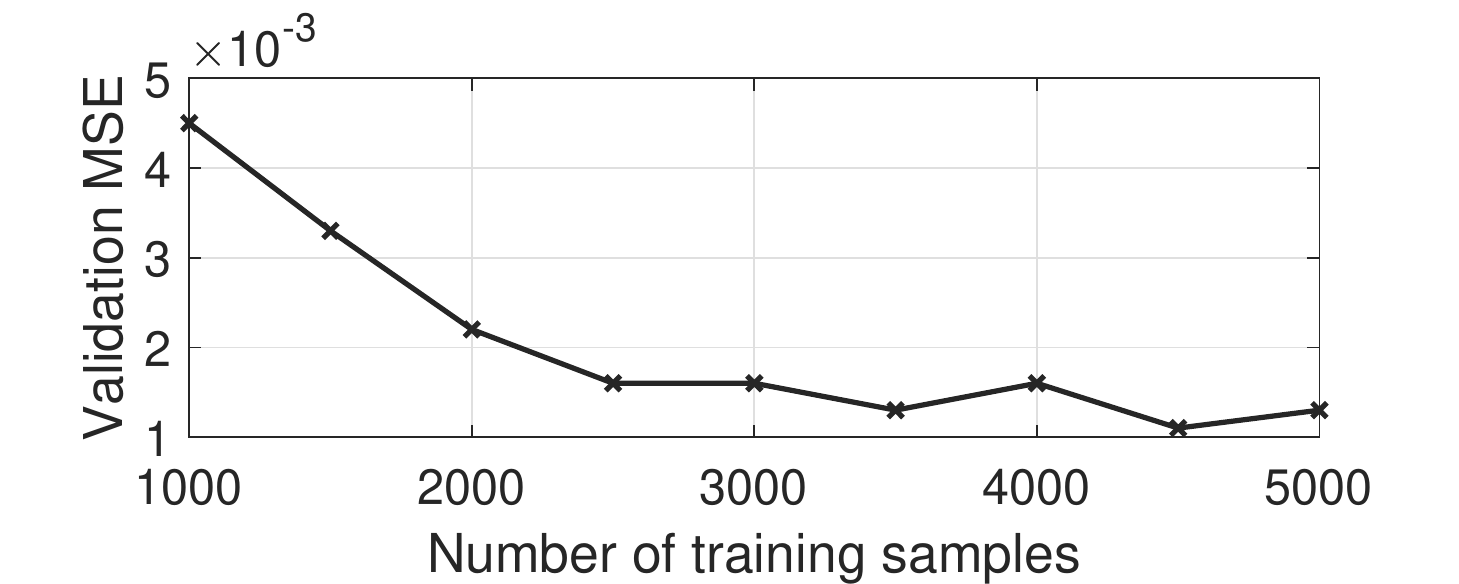}
\caption{Validation MSE as a function of number of training samples.}
\label{fig:validation_err}
\end{figure}

In order to train the CNN model, first, it is required to split the generated data-set into training, testing and validation sets. To realize the proper size of training data, training sets with different sizes from 1000 to 5000 samples are investigated and a validation set with 800 samples is used to evaluate the accuracy achieved with these training sizes. Validation MSE for different training sizes is shown in Fig. \ref{fig:validation_err}. Since there is not a considerable improvement in validation MSE error for training data-sets with more than 3500 samples, we choose this value as the training data size for all further analyses. 

Once the CNN model is trained, its performance accuracy is measured using a test data-set with 800 samples. To measure the accuracy, we apply regression metrics based on the the predicted and the true values of the pump setup for test data inputs. We choose coefficient of determination referred to as R-Squared ($R^2$) \cite{r_Square} as the measurement tool to interpret the goodness of fit of the trained network. $R^2$ takes values between 0 and 1 where a $R^2$ of 1 indicates that the variation of the regression predictions perfectly explain the variation of the true values. Contrarily, low values of $R^2$ indicate that the regression model outputs do not vary according to the true values. $R^2$ values for predicted pump powers in test data-set are shown in table \ref{r2 values}. It is indicated that $R^2$ values are higher than 0.90 for all pumps except $p_8$ with $R^2$ slightly lower than 0.90. These results confirm the very good performance of the CNN in modelling the mapping between the 2D profiles and their corresponding pump power values in the proposed second-order amplification scheme.

\begin{table}[h]
\caption{$R^2$ values of the pump power set for test data}
\label{r2 values}
\setlength{\tabcolsep}{2pt} 
\begin{tabularx}{\columnwidth}{|X | X | X | X| X| X| X|X|X|}
     \hline
    Pump  & $p_1$ & $p_2$ & $p_3$ & $p_4$ & $p_5$ & $p_6$ & $p_7$ & $p_8$ \\ \hline
    $R^2$  & $0.98$ & $0.93$ & $0.99$ & $0.97$ & $0.95$ & $0.96$ & $0.95$ & $0.86$\\ [3pt]

     \hline
\end{tabularx}

\end{table}

An alternative way to quantify the goodness of fit for the trained CNN model is to use the pump powers values predicted for the $i$th test sample as the input to the Raman solver in Eq. \ref{eq:Raman} to generate the corresponding 2D power profile $\hat {\textbf{P}}^i(f,z)$. The error is defined as the maximum absolute difference $E_{max}^{i}$ between the true 2D profile of $i$th test sample $ \textbf{P}^i(f,z)$ and the predicted 2D profile $\hat {\textbf{P}}^i(f,z)$, both defined in dBm scale, as the following:

\begin{equation}
E_{max}^{i} = \smash{\displaystyle\max_{z,f} |\textbf{P}^i(f,z) - \hat {\textbf{P}}^i(f,z)|}
\end{equation}

\begin{figure}
\centering
\includegraphics[width=8.3cm]{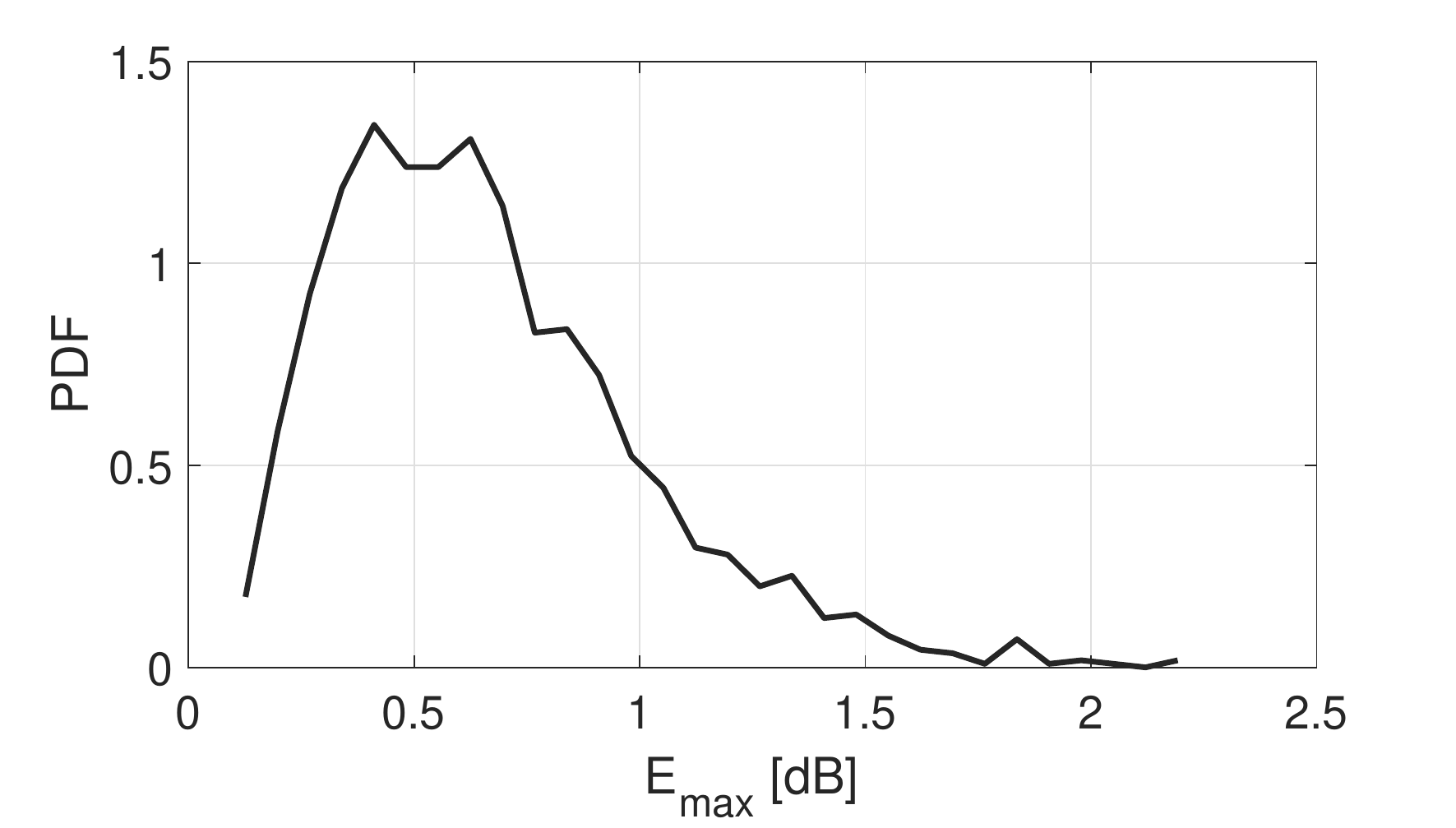}
\caption{PDF of $E_{max}$ for the test data.}
\label{fig:erro max}
\end{figure}

Probability density function (PDF) of the $E_{max}$ for test data-set is shown in Fig. \ref{fig:erro max}. The low mean $\mu = 0.62$ dB and the standard deviation $\sigma = 0.33$ dB values for $E_{max}^{i}$ also assert the good performance of the proposed CNN model for the pump powers values prediction. 

\subsection{Evaluation of the CNN-assisted DE framework on designing specific profiles}

Signal power profiles defined over the spectral and spatial domains such as 2D flat and symmetric with respect to the middle point of the span, are the two mostly addressed profiles in the literature due to their attractive properties \cite{Quasi-lossless1, Ania-Castanon:04, tan2018distributed, Rosa:15}. The CNN-assisted DE framework, as discussed in section II, aims to minimize a pre-defined cost value $J$. Mainly, the value of $J$ depends on some requirements on the shape of the target 2D power profile. In the following, we explain the formulation of the cost function for 2D flat and symmetric power profiles. Once the cost function is defined, we present the results of applying the CNN-assisted DE technique discussed in the previous section.

\subsubsection{\textbf{2D flat power evolution}}We follow the formulation presented in the literature mainly for single channel and extend it to our wide-band with multi-channel (whole C-band) analysis \cite{PhysRevLett.101.123903, Quasi-lossless1}. Additionally, we formulate the objective functions as a function of the pump powers values $\textbf{p}_{pump}$ to fit properly in the proposed optimization framework. 

We approach the 2D flat profiles design problem with three cost functions. Each of them applies a specific assessment quantity on the shape of the 2D profile. Considering this relation, the overall optimization problem is to find $\textbf{p}_{pump}^{*}$ which jointly minimizes the three costs defined consecutively. 

Assuming the signal power level defined in dBm scale, the main objective in a 2D flat profile design is to minimize the maximum power excursion in frequency and along the fiber span $J_0$ (in dB scale). Maximum power excursion is defined as the difference between the maximum and the minimum of the signal power level in spectral-spatial plane:
\begin{multline}
\label{eq:F0}
 J_0(\textbf{p}_{pump}) = 
 \\ \smash{\displaystyle\max_{f,z}(\textbf{P}(f,z|\textbf{p}_{pump}))} -  \smash{\displaystyle\min_{f,z}(\textbf{P}(f,z|\textbf{p}_{pump}))}
\end{multline}
As detailed in section I, one of the aspirations in a wide-band amplification system is to achieve signal power with minimum spectrum excursion at the receiver side \cite{8894395, deMoura:20, 9244561}. In this paper, we generalize its definition by aiming to minimize the maximum spectrum excursion among all distance points in the fiber. We refer to this cost function as spectrum excursion $J_1$ defined as the following:
\begin{multline}
\label{eq:F1}
 J_1(\textbf{p}_{pump}) =\\ \smash{\displaystyle\max_{z}[\smash{\displaystyle\max_{f}(\textbf{P}(f,z|\textbf{p}_{pump}))} - \smash{\displaystyle\min_{f}(\textbf{P}(f,z|\textbf{p}_{pump}))}]} 
\end{multline}
Additionally, in the experimental setups presented in the literature for flat power evolution design \cite{PhysRevLett.101.123903, Quasi-lossless1}, the pump powers values are tuned empirically to provide a 0 dB gain between the transmitter and the receiver. The 0 dB gain $J_2$ between the transmitter and the receiver is considered here as the third cost value with the following formulation:
\begin{multline}
\label{eq:F2}
 J_2(\textbf{p}_{pump}) =\\ \smash{\displaystyle\max_{f}|\textbf{P}(f,z=L|\textbf{p}_{pump}) - \textbf{P}(f,z=0|\textbf{p}_{pump})}|
\end{multline}
where $L$ is the fiber length and $|.|$ is the absolute value operator.

Considering these objectives, we define a multi-objective optimization problem for 2D flat target profile aiming to find the best $\textbf{p}_{pump}$ set. Due to the complexity of each cost function, it is challenging to realize whether the minimum value of the three costs occur at the same time or not. Therefore, to make the optimization process simpler and also to be able to control the impact of each objective, we make an approximation by converting it into a classical weighted-sum optimization. Approaching this, each objective is multiplied by a weight, defined as a hyper-parameter, and it is added to the other objectives as the following: 

\begin{align}
\label{eq:so}
&\textbf{p}_{pump}^{*} = \smash{\displaystyle\underset{\textbf{p}_{pump}}{\arg\min} \sum_{i=0}^{2}{m_iJ_i(\textbf{p}_{pump})}} = \textbf{m}\cdot\textbf{J}^T(\textbf{p}_{pump}),  \\
\notag  \\
 &\text{such that   }\textbf{p}_{LB} \leq \textbf{p}_{pump} \leq \textbf{p}_{UB}, \;  m_i>0, \;  \sum_{i=0}^{2}{m_i}=1  \notag \\
 \notag
\end{align}

where $\textbf{m} = [m_0, m_1, m_2]$ is a hyper-parameter vector of weights, and each element of it controls the impact of the corresponding cost in the objective vector $\textbf{J}(\textbf{p}_{pump}) = [J_0(\textbf{p}_{pump}), J_1(\textbf{p}_{pump}), J_2(\textbf{p}_{pump})]$, and $T$ is the transpose operator. An advantage of the DE as a gradient-free optimization technique is that we can target problems with multiple objectives which we cannot solve easily with an ML-based inverse model like a CNN. 



\begin{figure*}[hbt!]
     \centering
     \begin{subfigure}
         \centering
         \includegraphics[width=0.85\columnwidth]{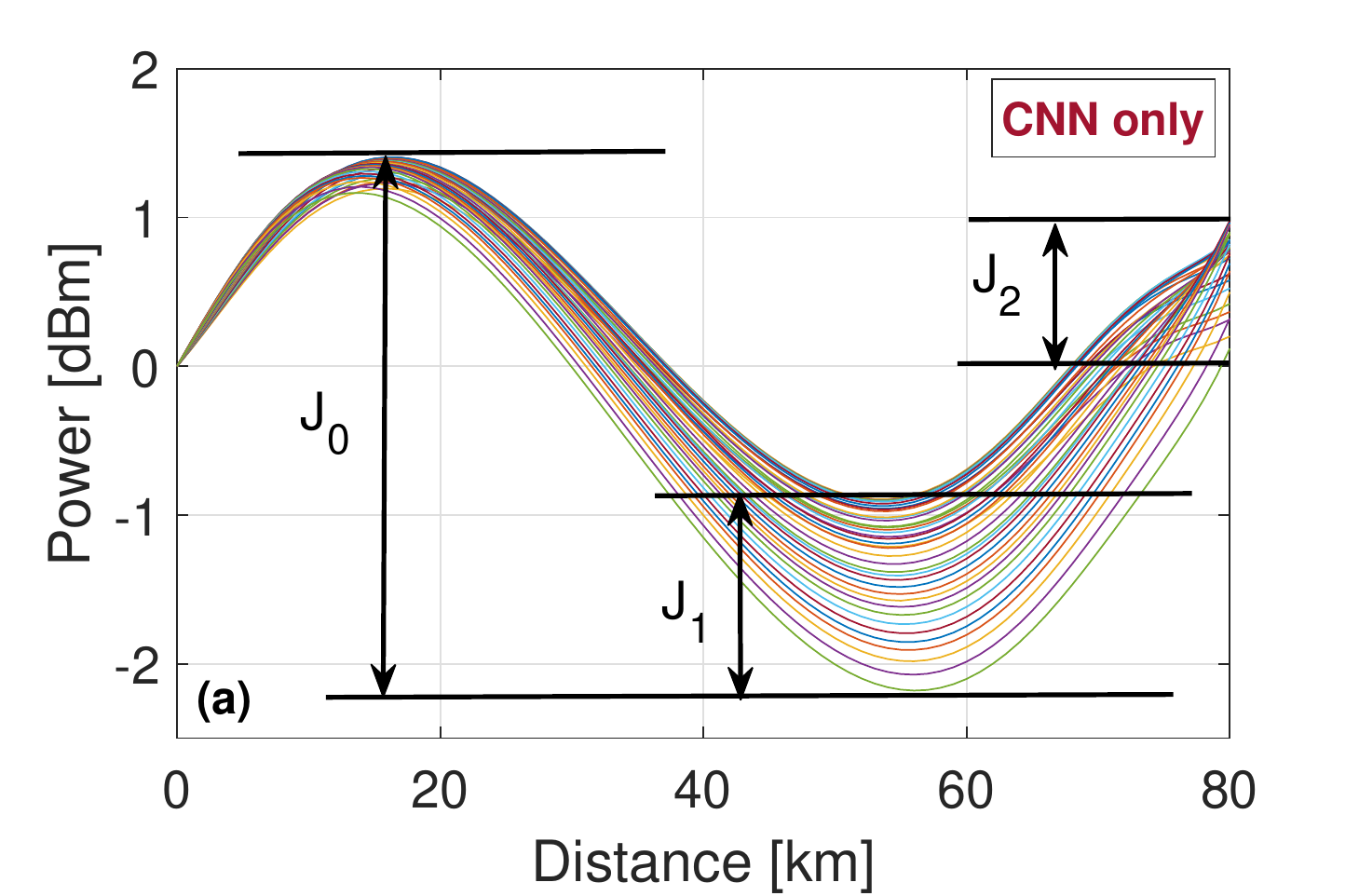}
         \label{fig:cnn out}
     \end{subfigure}
     \begin{subfigure}
         \centering
         \includegraphics[width=0.85\columnwidth]{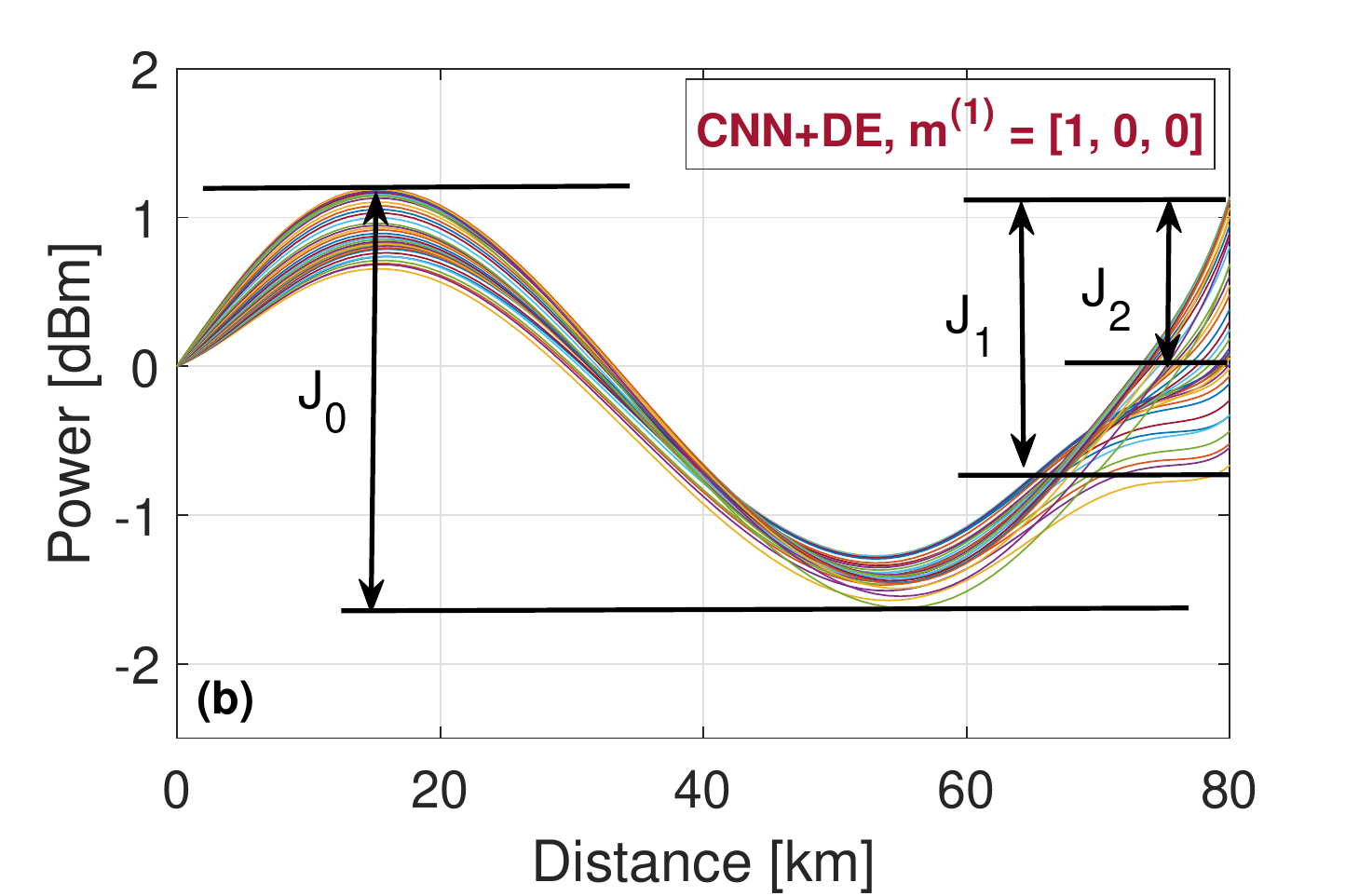}
         \label{fig:m 1 0 0}
     \end{subfigure}
     \begin{subfigure}
         \centering
         \includegraphics[width=0.86\columnwidth]{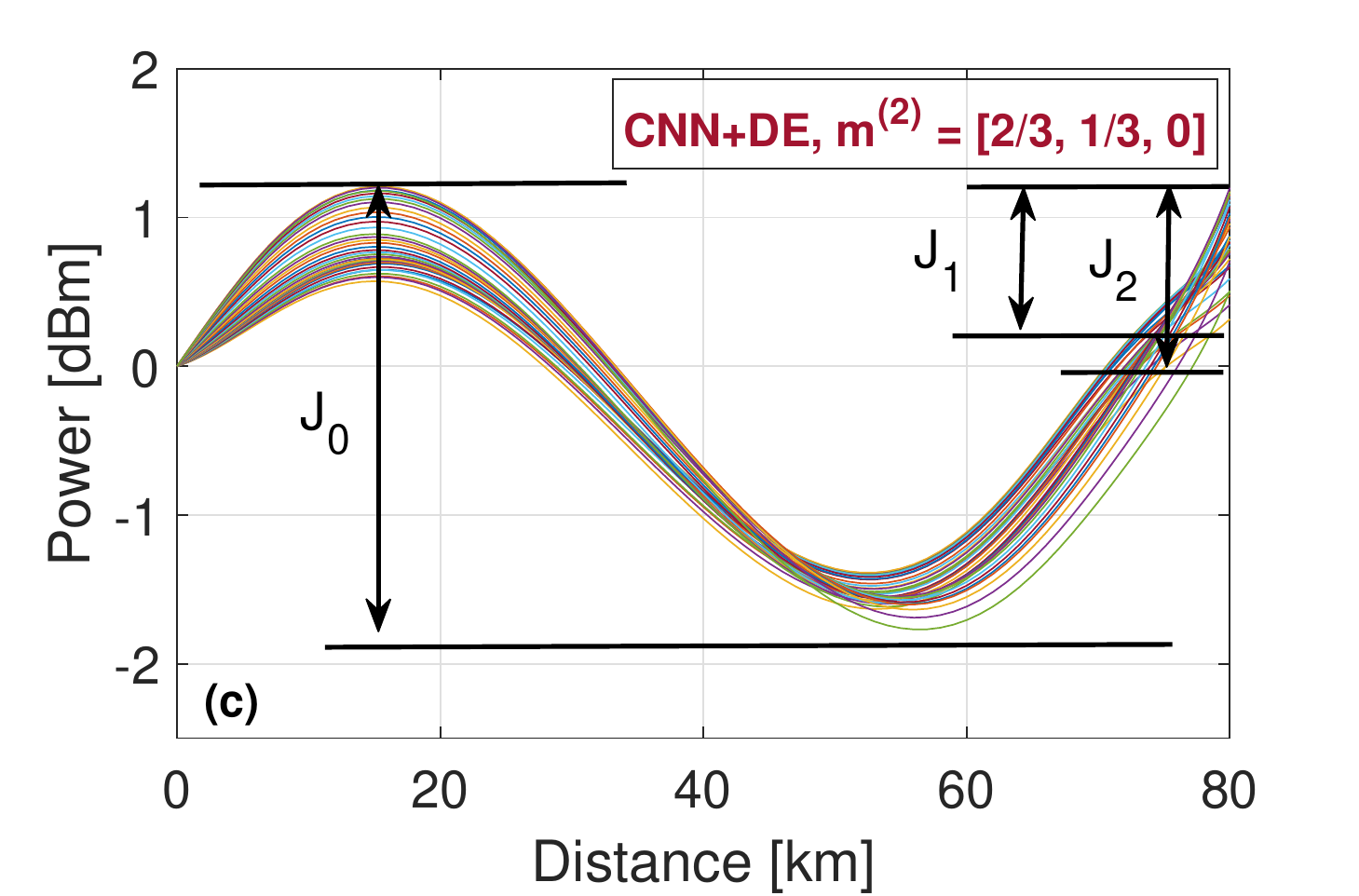}
         \label{fig:m 1 1/2 0}
     \end{subfigure}
          \begin{subfigure}
         \centering
         \includegraphics[width=0.85\columnwidth]{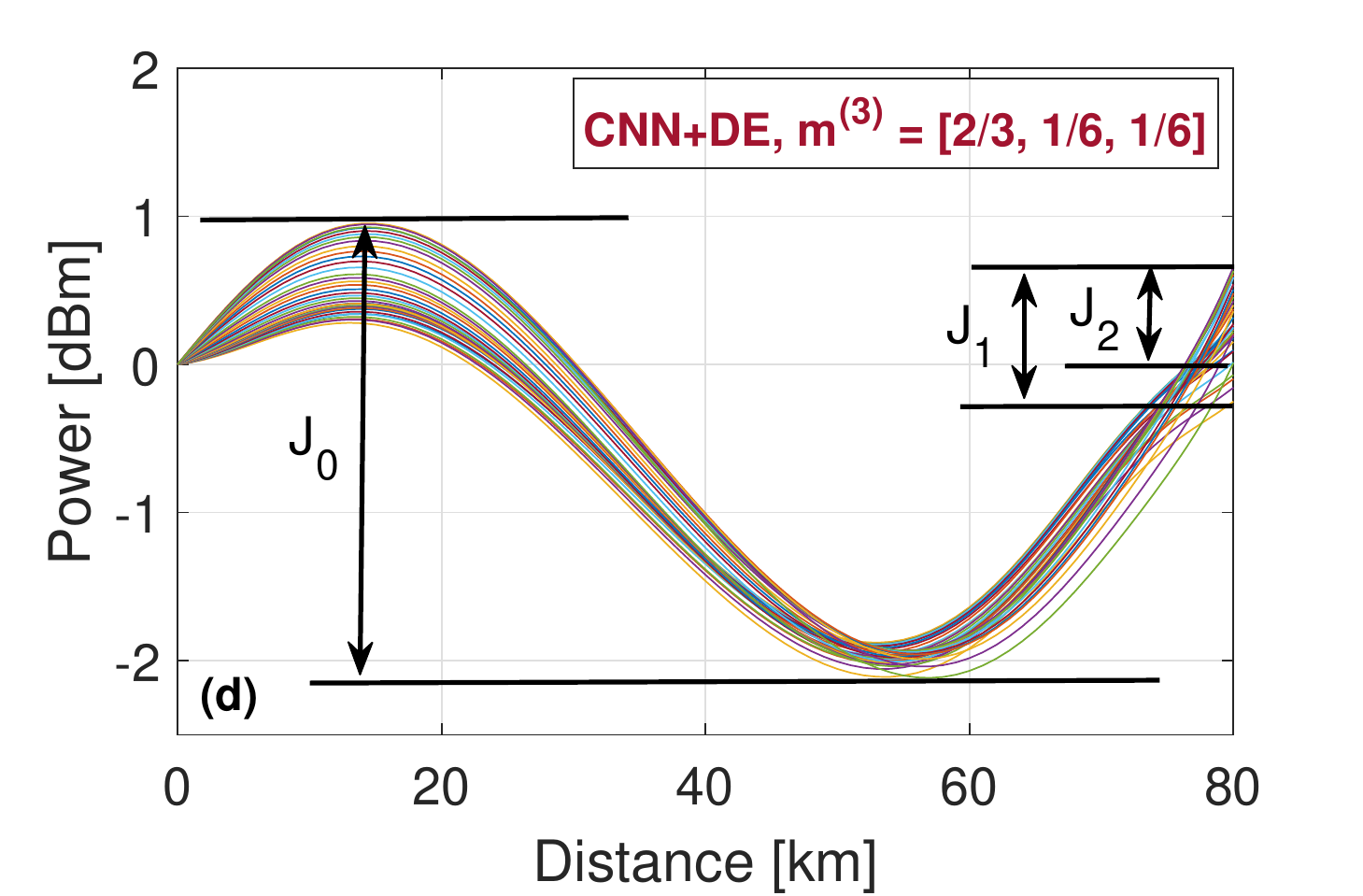}
         \label{fig:m 1 1/4 1/4}
     \end{subfigure}
        \caption{Resulting 2D power evolution profiles for a 2D flat target profile with their corresponding objectives $[J_0, J_1, J_2]$. (a) CNN without the fine-tuning process, (b) CNN-assisted DE with $\textbf{m}^{(1)} = [1, 0, 0]$, (c) CNN-assisted DE with $\textbf{m}^{(2)} = [2/3, 1/3, 0]$, (d) CNN-assisted DE with $\textbf{m}^{(3)} = [2/3, 1/6, 1/6]$.
        }
        \label{fig:flatness results}
\end{figure*}
 
After defining the cost function for the 2D flat profile, we apply the framework presented in Fig. \ref{fig:mainmethod} to find the corresponding $\textbf{p}_{pump}^{*}$. In this approach, first, we create a 2D power profile with 0 dBm power level as the 2D flat input to the CNN network to evaluate its performance in finding the corresponding set of pump powers values. The predicted pump powers by the CNN model for the 2D flat input profile are reported in table \ref{power values}, in the column labeled with CNN. 

Following the framework in Fig. \ref{fig:mainmethod}, the pump power values predicted by the CNN are used to initialize the fine-tuning process performed by the DE algorithm.  Meanwhile, as presented in algorithm \ref{alg:cap}, DE has a set of hyper-parameters, which needs to be specified. A possible way to find the best set of hyper-parameters is to do a grid-search by setting the combination of all possible values of them. However, grid-search on finding the best set of hyper-parameters will make the problem complex and time consuming. In order to reduce the complexity, we set some of the DE hyper-parameters values such as the crossover probability and the mutation factor, same as the standard values $CR = 0.5$ and $F = 0.8$, normally used for other applications \cite{5601760}. The population size $N_p$ is set to 30 which is three times more than the dimension of the search space, recommended as a plausible value in the DE literature \cite{5601760}. Additionally, the maximum number of cost function evaluation is set as $MaxEv = 1000$. 

Regarding $\boldsymbol\Delta_p$, we set the same and fixed scaling factor value for the first-order pumps, and also the same value for the second-order pumps. Normally, second order pumps have relatively high value and due to Raman pump-to-pump interaction, they have remarkable impact on the first order pump. Therefore, in $\boldsymbol\Delta_p$, we consider lower corresponding scaling factor for the second-order pumps compared to the first-order ones. Considering the higher predicted value of $p_5$ in comparison with the value predicted for $p_1$, we consider 1.4 [W] as the maximum possible value and consequently $\Delta_1 = \Delta_5 = 0.35 $. Addressing the first-order pumps, based on the highest value which is predicted for $p_3$, we consider more deviation to first-order pumps by setting the corresponding scaling factor value to 0.5 resulting in $\boldsymbol\Delta_p = [0.35, 0.5, 0.5, 0.5 , 0.35 , 0.5, 0.5, 0.5]$ (0.35 scaling value for the second-order pumps and 0.5 for the first-order pumps).




\begin{table}[h]
\caption{Predicted pump power values by the CNN and CNN-assisted DE with different weights for 2D flat input power profile}
\centering
\label{power values}
\begin{tabular}{|l||  p{0.15\linewidth}| p{0.15\linewidth}| p{0.15\linewidth}| p{0.15\linewidth}|}\hline

    \hline
    \backslashbox[17mm]{Pump}{model}
    & CNN only& CNN+DE with $\textbf{m}^{(1)}$& CNN+DE with $\textbf{m}^{(2)}$ & CNN+DE with $\textbf{m}^{(3)}$\\ \hline\hline
    $p_1[mW]$ & 330 & 430 & 440 & 450\\ [3pt]\hline
    $p_2[mW]$ & 33 & 45 & 49 & 47 \\[3pt] \hline
    $p_3[mW]$ & 145 & 98 & 90 & 76 \\[3pt] \hline
    $p_4[mW]$ & 12 & 12 & 13 & 14 \\ [3pt]\hline
    $p_5[mW]$ & 1030 & 1150 & 1060 & 990 \\[3pt] \hline
    $p_6[mW]$ & 12 & 8 & 6 & 6 \\ [3pt]\hline
    $p_7[mW]$ & 19 & 12 & 18 & 21 \\ [3pt]\hline
    $p_8[mW]$ & 43 & 24 & 63 & 63 \\[3pt] \hline

\end{tabular}
\end{table}

According to Eq. \ref{eq:so}, impact of the each cost $J_i, i = 0, 1, 2$ on the overall optimization problem is controlled by the corresponding weight value $m_i$. This approach provides the flexibility to target desired objectives with different weights. We consider the minimum power excursion defined in Eq. \ref{eq:F0} as the objective with the highest impact, and accordingly, propose three scenarios with different values of $\textbf{m} = [m_0, m_1, m_2]$. First, we set $\textbf{m}$ as $\textbf{m}^{(1)} = [1, 0, 0]$ where the objective is only the minimization of the power excursion $J_0$, secondly $\textbf{m}^{(2)} = [2/3, 1/3, 0]$ where the objective is to minimize both the $J_0$ and the spectrum excursion $J_1$ with considering half impact than $J_0$. Finally, we set $\textbf{m}^{(3)} = [2/3, 1/6, 1/6]$ where the objective is to minimize $J_0$ , $J_1$ and 0 dB gain variation $J_2$ giving equal impact to $J_1$ and $J_2$, each one with quarter impact of $J_0$. Table \ref{power values} shows the pump power values achieved for the CNN-assisted DE scenarios with the aforementioned $\textbf{m}$ values.

To have an insight on the performance of the proposed 2D flat power profile design problem, we generate the power evolution profiles resulted from the CNN and also the CNN-assisted DE framework for different $\textbf{m}$ values. Regarding this, the power evolution profiles over the distance consisting of 40 channels for the proposed scenarios are shown in Fig. \ref{fig:flatness results}.

Fig. \ref{fig:flatness results} (a) shows the power evolution for the pump parameters predicted by the CNN, while Fig. \ref{fig:flatness results} (b), (c) and (d) show the CNN-assisted DE optimization for $\textbf{m}^{(1)}$, $\textbf{m}^{(2)}$ and $\textbf{m}^{(3)}$ values, respectively. The cost functions values $J_0$, $J_1$ and $J_2$ are specified on each one of the sub-figures and the corresponding values are reported in table \ref{objectives values}. It is shown that the minimum power excursion is achieved for the CNN-assisted DE with $\textbf{m}^{(1)}$ value which is $2.81$ dB. However, this case has relatively high spectrum excursion $J_1$ and also high 0 dB gain $J_2$ variation values. 

Considering the same framework with $\textbf{m}^{(2)}$, the spectrum excursion improves almost $0.9$ dB while the power excursion increases less than $0.2$ dB. On the other hand, the 0 dB gain variation for both cases is still higher than $1$ dB. By involving 0 dB gain variation in the fine-tuning process by using $\textbf{m}^{(3)}$, it improves more than $0.5$ dB while we observe an increase in the power excursion value by less than $0.1$ dB. It has been illustrated that the position of the maximum spectrum excursion $J_1$ occurs at the end of the fiber link except for the output of the CNN model which is found at about the distance of 55 km (Fig. \ref{fig:flatness results} (a)). 

\begin{table}[ht]
\caption{Cost function values achieved by CNN and CNN-assisted DE framework for a 2D flat input profile}
\centering
\begin{tabular}{|l|| p{0.15\linewidth}| p{0.16\linewidth}| p{0.16\linewidth}| p{0.16\linewidth}|}\hline

    \hline
    \backslashbox[15mm]{Cost}{model}
    & CNN only & CNN+DE with $\textbf{m}^{(1)}$&CNN+DE with $\textbf{m}^{(2)}$&CNN+DE with $\textbf{m}^{(3)}$\\ \hline\hline
    $J_0$&3.58 dB& \textbf{2.81} dB & 2.97 dB & 3.06 dB \\ \hline
    $J_1$ &1.48 dB& 1.80 dB & \textbf{0.88} dB & 0.90 dB \\ \hline
    $J_2$ &0.97 dB& 1.14 dB & 1.2 dB & \textbf{0.65 dB} \\ \hline

\end{tabular}
\label{objectives values}
\end{table}

 \begin{figure}[hbt!]
    \centering
    \includegraphics[width = 8.5cm]{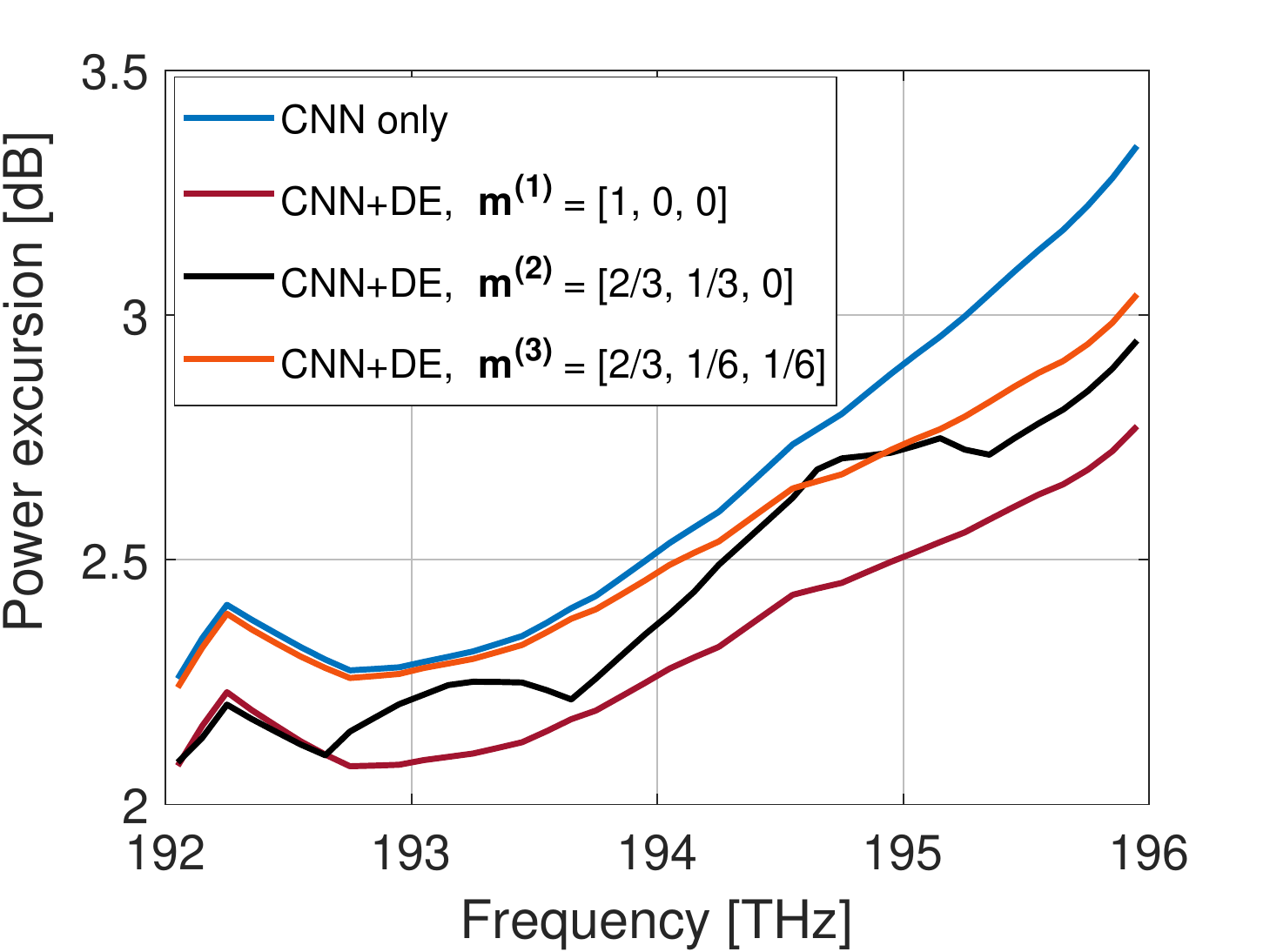}
    \caption{Power excursion value computed as a function of frequency, shown for  different pump power adjustment scenarios.}
    \label{fig:flatness per channel}
\end{figure}

In addition to the wide-band analysis of power excursion in a 2D profile, we also analyze the power excursion as a function of frequency. For the predicted $\textbf{p}_{pump}^*$ in each proposed 2D profile design scenario, the difference between the maximum and the minimum value of signal power level at each frequency value over the all distances is shown in Fig. \ref{fig:flatness per channel}.

It is illustrated that generally, the power excursion value is increasing with frequency for CNN and also all the CNN-assisted DE scenarios. Additionally, the minimum power excursion which is less than $2.4$ dB, is achieved almost at 192.7 THz. The maximum value for all scenarios is measured at 196 THz which is less than the corresponding power excursion $J_0$ measured based on Eq. \ref{eq:F0} for the whole 2D frequency-distance plane.

\begin{figure}[hbt!]
    \centering
    \includegraphics[width = 8.5cm]{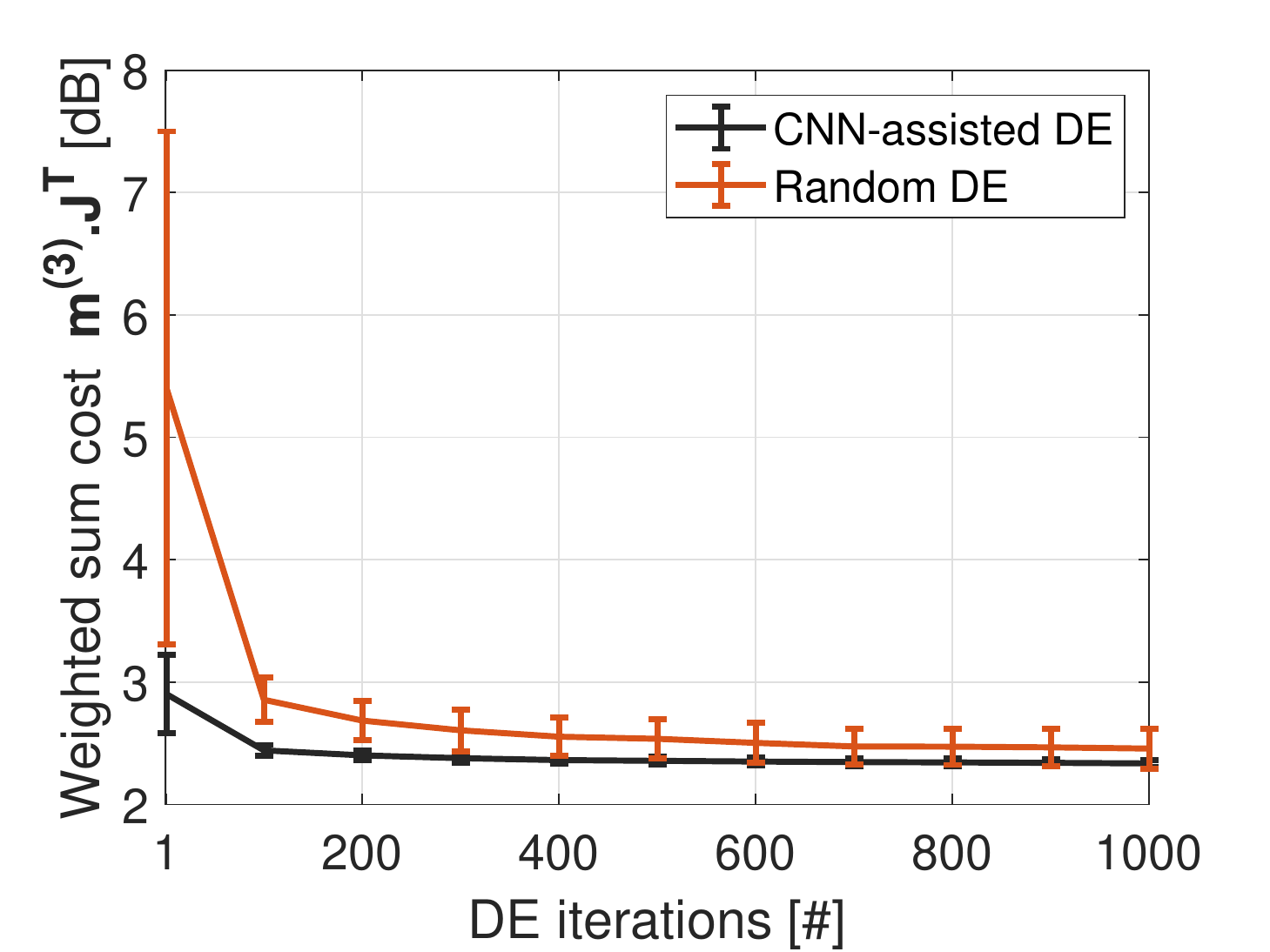}
    \caption{Average cost (plot line) and standard deviation (error bars) for the CNN-assisted DE and the DE optimization with random initialization for flatness analysis as a function of number of DE iterations.}
     \label{flat error bar}
\end{figure}

In order to test the advantage provided by the CNN, we consider the case in which the optimizer is left free to span the whole pump power space, i.e. setting $\textbf{p}_{LB}$ and $\textbf{p}_{UB}$ to the minimum and the maximum pump power values specified in table \ref{pump ranges}, respectively. In this case, referred to as "random DE", the DE population will be initialized in a wide space of pump power values. We run the algorithm for ten trials to see how is the average performance over the number of DE iterations for different randomly initialized populations. Additionally, we also perform ten trials for the CNN-assisted DE to compare its average performance with random DE. The objective of all trials for the both approaches is to solve Eq. \ref{eq:so} with $\textbf{m}^{(3)}$. 

In Fig. \ref{flat error bar}, the error bars of these two approaches are compared. The line plot determines the average error and the vertical error bar at each point shows one standard deviation above and below the average error. In average, the CNN-assisted DE converges to a lower minimum value of cost compared to the random DE. Furthermore, error bars show that the standard deviation of the cost value $\textbf{m}^{(3)}\cdot \textbf{J}^T$ at each iteration is relatively smaller for the CNN-assisted DE, indicating more reliability of using the CNN for initialization of the population by converging to the same minimum point for different trials.

\subsubsection{\textbf{2D symmetric spatial power evolution}}Symmetric power evolution in distance has been addressed so far mostly for single-channel narrow-band amplification scenarios \cite{Rosa:15, Rosa:2015}. The main purpose for symmetric power evolution design is to numerically or experimentally minimize the asymmetry factor $A(f)$ at a desired frequency $f_0$ defined as the following \cite{Rosa:15}:

\begin{equation}
\label{eq:sym1}
 A(f_0) = \frac{\int_{0}^{L/2} |\textbf{P}(f_0, z)-\textbf{P}(f_0, L-z)| dz}{\int_{0}^{L/2} \textbf{P}(f_0, z) dz}
\end{equation}
 
where the power profile $\textbf{P}(f, z)$ is defined in linear scale [mW]. Generalizing the minimization model to a wide-band scenario, we define the proposed cost value $J$ as the maximum asymmetry over all frequency channels existing in the bandwidth:
 
\begin{equation}
\label{eq:asym}
 J = \max_{f}(A(f))
\end{equation}

 in which $A(f)$ and consequently $J$ value, can be formulated as a function of the pump power values as the following:
 
\begin{multline}
\label{eq:asym}
 J(\textbf{p}_{pump}) =  \\
 \max_{f}[\frac{\int_{0}^{L/2} |\textbf{P}(f_0, z|\textbf{p}_{pump})-\textbf{P}(f_0, L-z|\textbf{p}_{pump})| dz}{\int_{0}^{L/2} \textbf{P}(f_0, z|\textbf{p}_{pump}) dz}]
\end{multline}
 
 and the optimization problem is formulated as: 
 
 \begin{align}
 \label{eq:sym1_opt}
 &\textbf{p}_{pump}^{*} = \smash{\displaystyle\underset{\textbf{p}_{pump}}{\arg\min} \; J(\textbf{p}_{pump}) }\\
  \notag \\
 &\text{such that \:   }\textbf{p}_{LB} \leq \textbf{p}_{pump} \leq \textbf{p}_{UB} \notag \\
 \notag
\end{align}

Similar to the approach for flat power evolution design, after defining the cost value for this target profile, we apply the proposed CNN-assisted DE framework illustrated in Fig. \ref{fig:mainmethod}.

The initial step in running the CNN-assisted diagram for designing a 2D symmetric power evolution is to provide a target 2D symmetric profile input for the CNN model. Unlike the flat power profile there are many 2D symmetric profiles which can be used as the input to the CNN model. In this paper, we have considered the second half-period of a sinusoidal signal as the symmetric power profile target for all channels. More specifically, we have formed a 2D symmetric power evolution, defined in logarithmic scale [dBm], with the following formulation:

\begin{equation}
\label{eq:sin_wave}
\textbf{P}^t(f,z) = 4sin(\pi z/L + \pi),\: \forall f, \: 0<z<L
\end{equation}

\begin{figure*}[hbt!]
     
     \begin{subfigure}
         \centering
         \includegraphics[width=0.7\columnwidth]{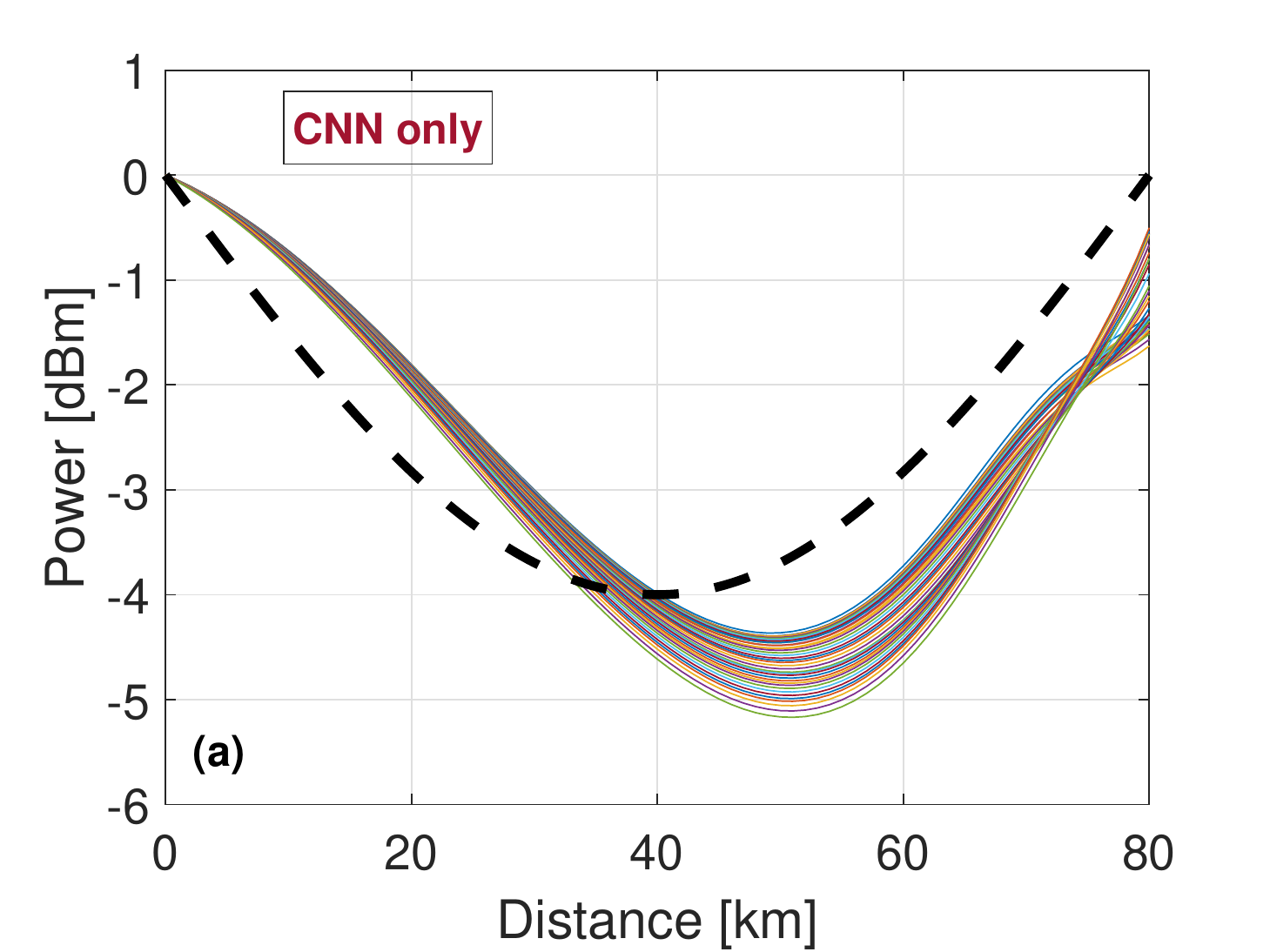}
         \label{fig:y equals x}
     \end{subfigure}
     \begin{subfigure}
         \centering
         \includegraphics[width=0.7\columnwidth]{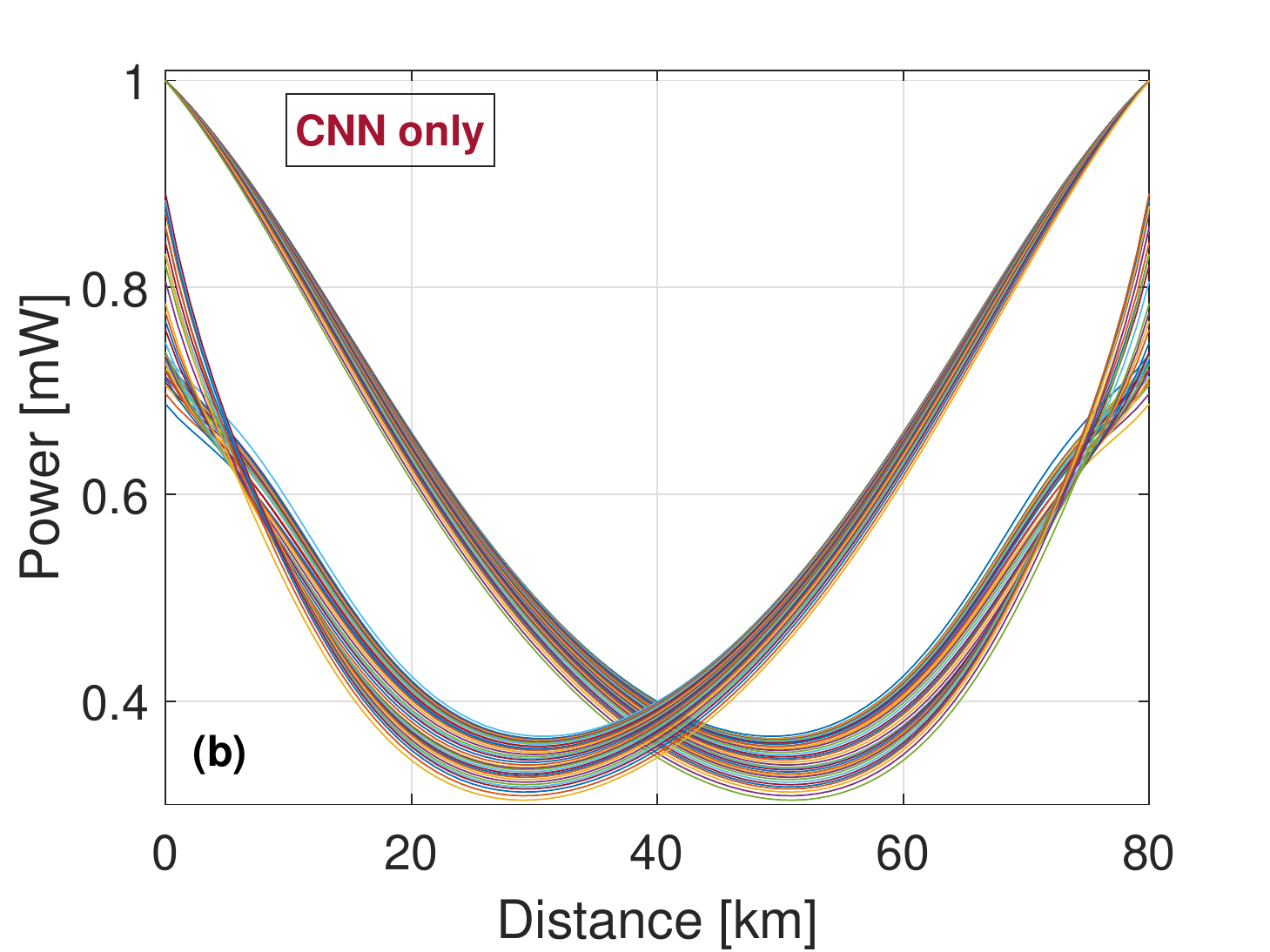}
         \label{fig:three sin x}
     \end{subfigure}
     \begin{subfigure}
         \centering
         \includegraphics[width=0.7\columnwidth]{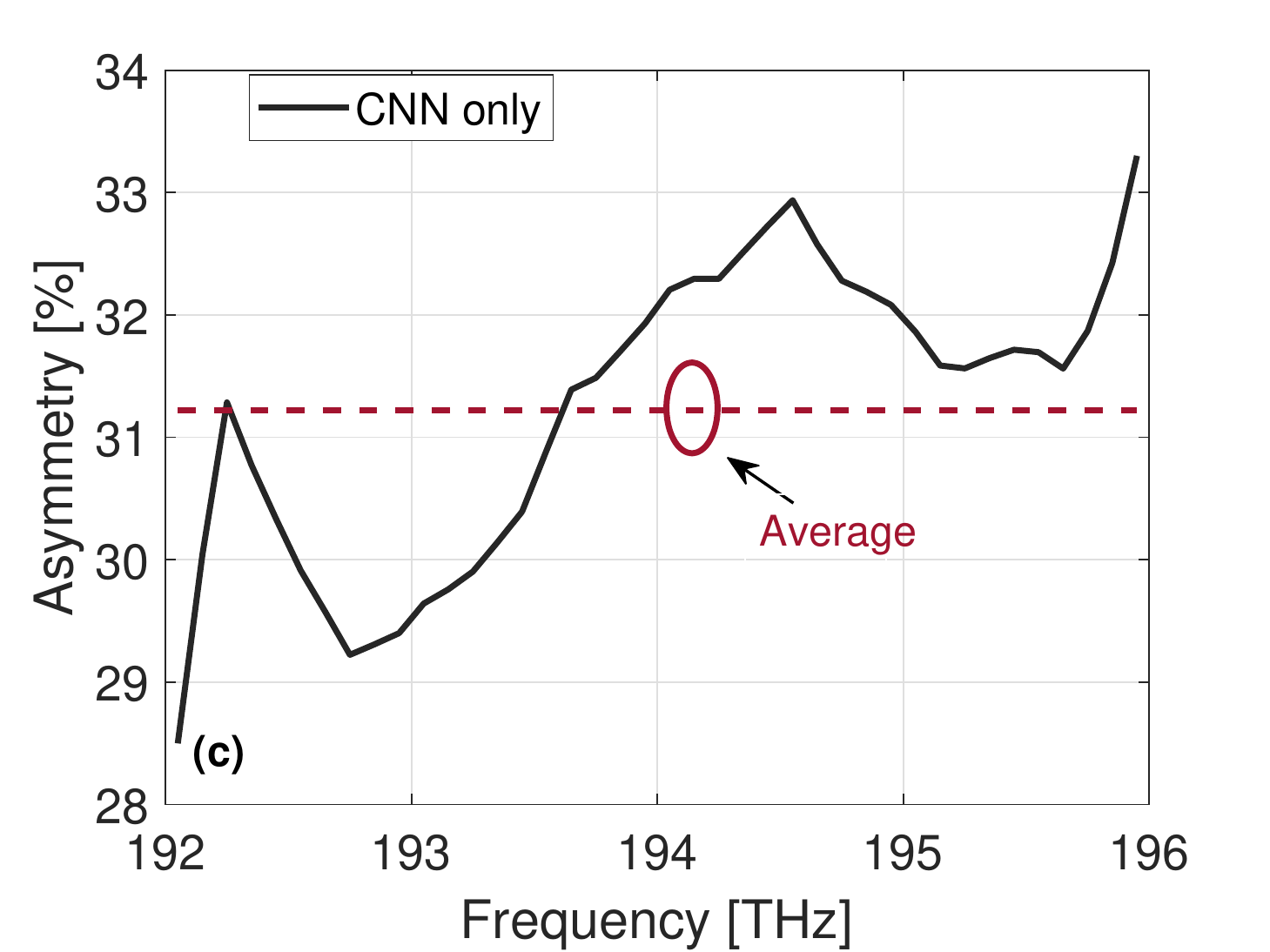}
         \label{fig:five over x}
     \end{subfigure}
     \\
     \begin{subfigure}
         \centering
        \includegraphics[width=0.7\columnwidth]{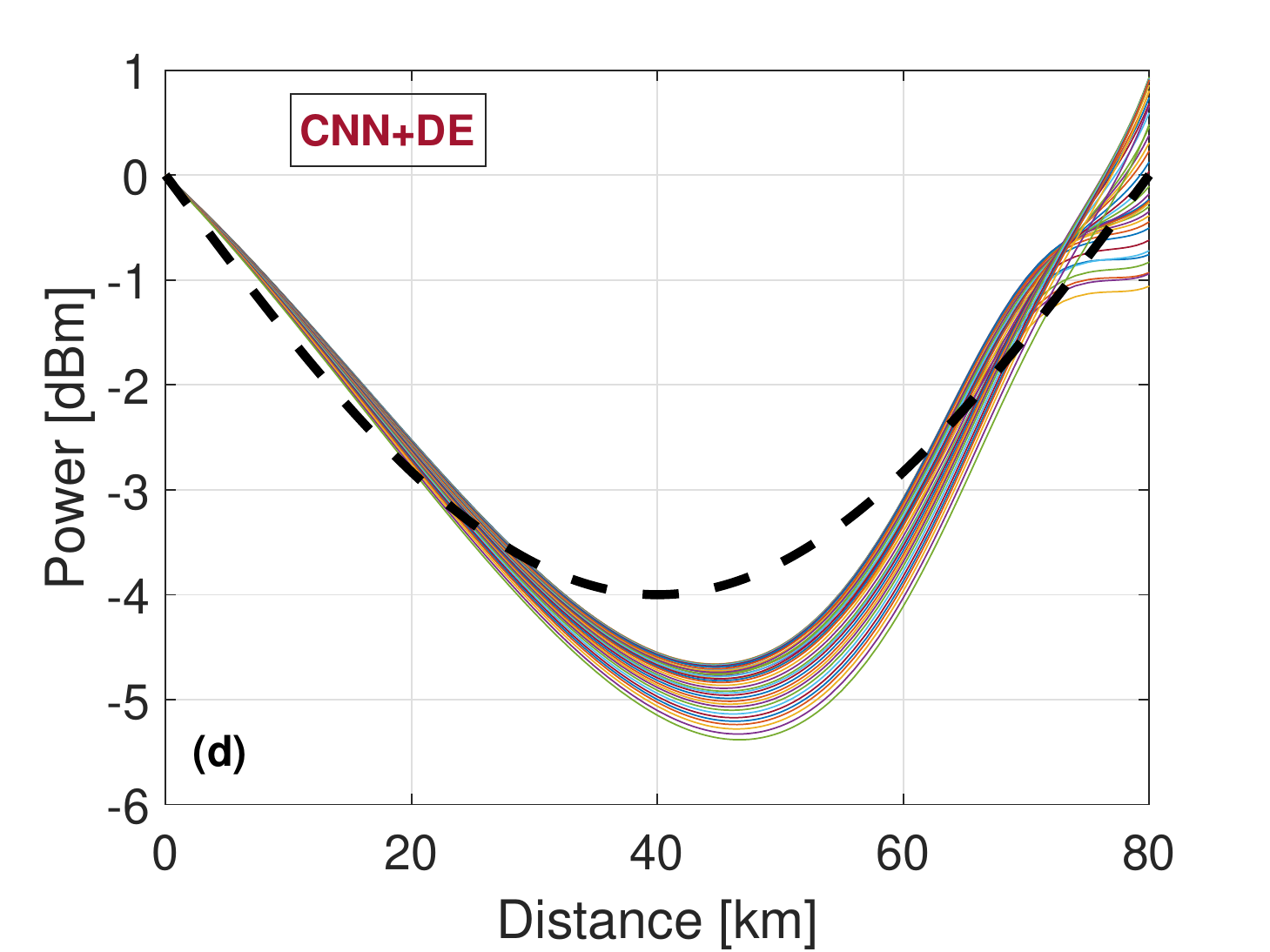}
         \label{fig:five over x}
     \end{subfigure}
     \begin{subfigure}
         \centering
         \includegraphics[width=0.7\columnwidth]{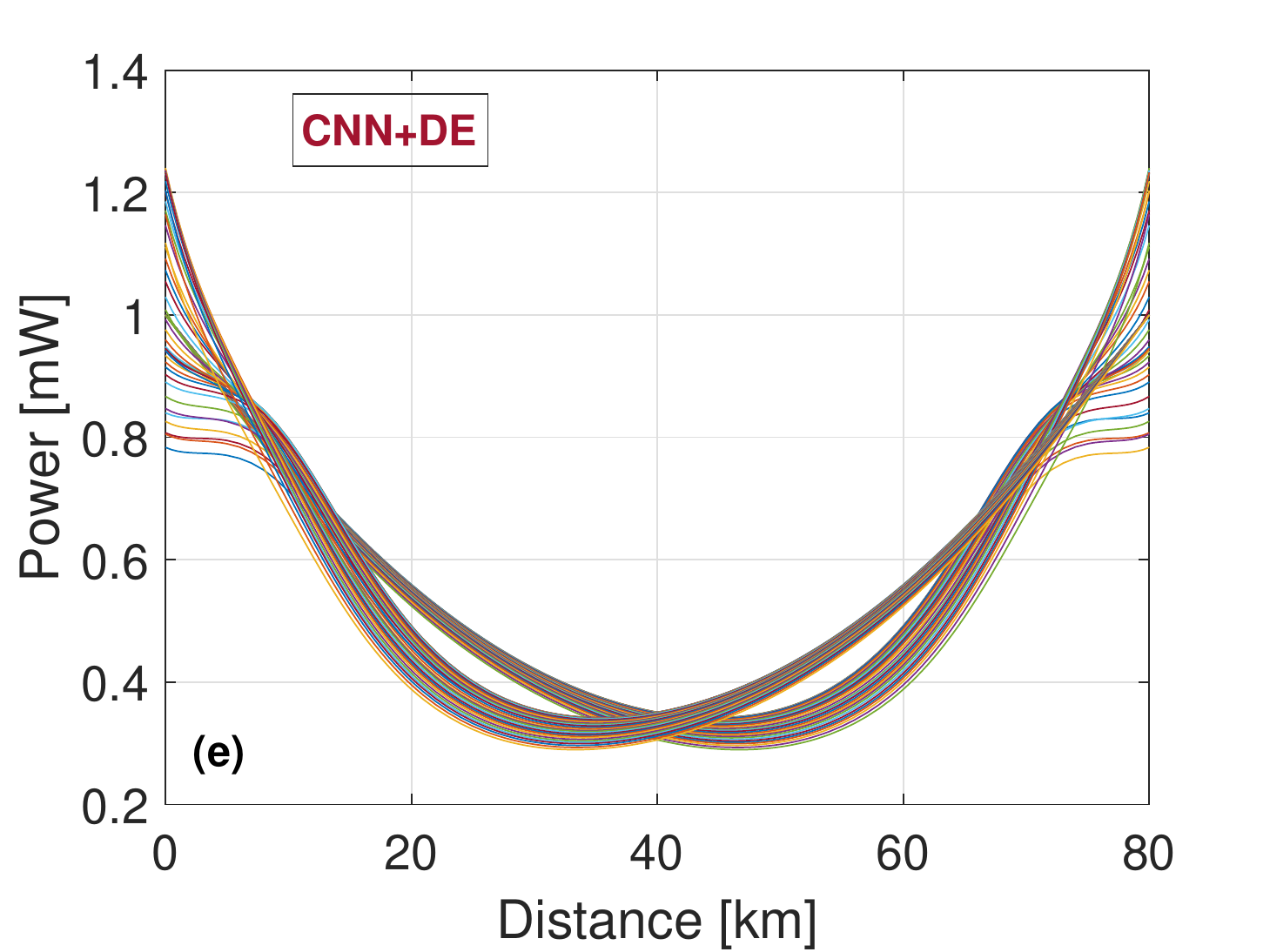}
         \label{fig:five over x}
     \end{subfigure}
     \begin{subfigure}
         \centering
         \includegraphics[width=0.7\columnwidth]{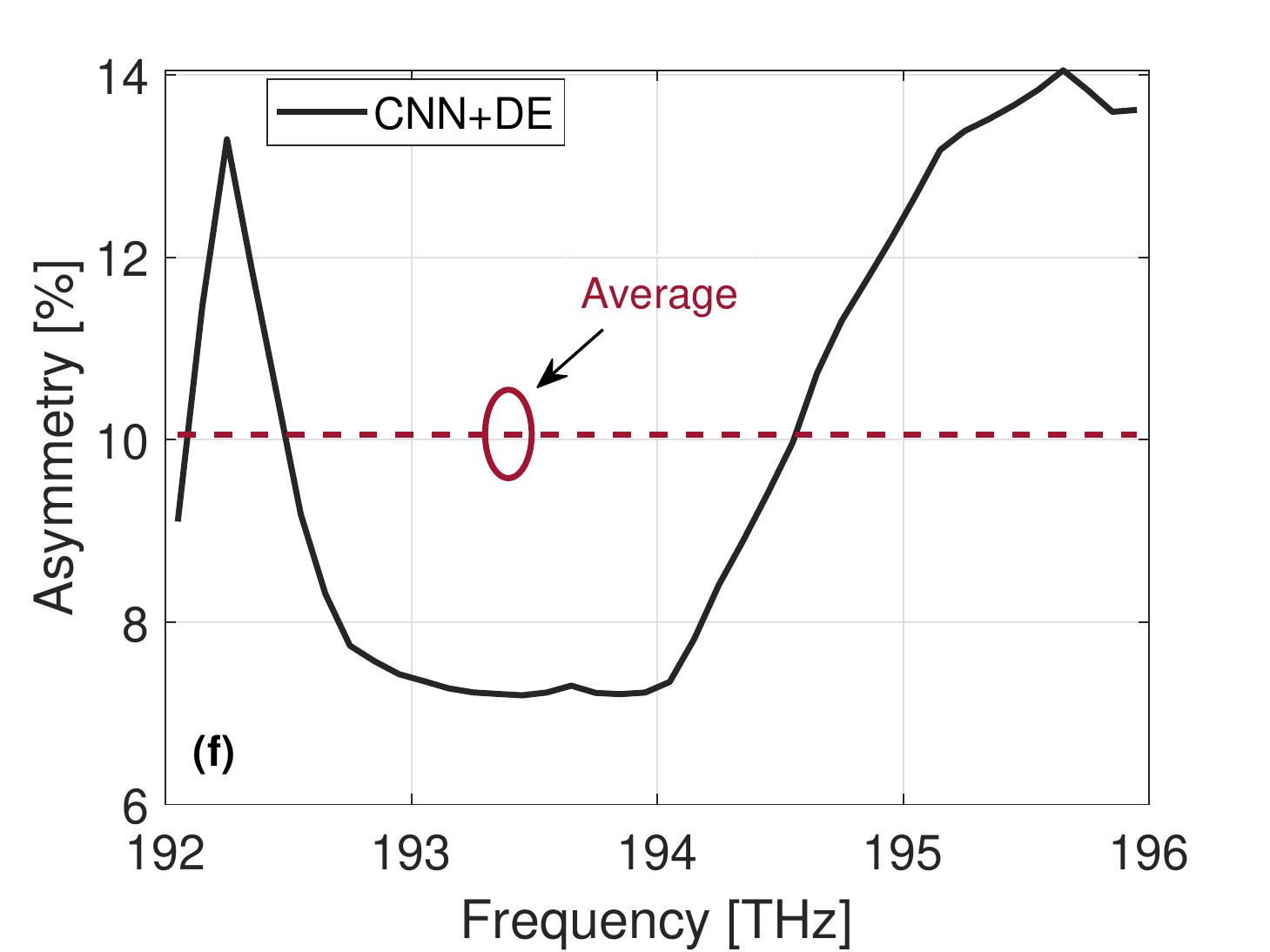}
         \label{fig:five over x}
     \end{subfigure}
        \caption{Results of symmetric power evolution design. (a) The power evolution generated based on the pump power values predicted by the CNN model (dashed black curve is the symmetric target as a function of distance for all channels based on Eq. \ref{eq:sin_wave}), (b) The power evolution and its reversed version in linear (mW) scale for the CNN, (c) The asymmetry in frequency for the CNN, (d) The power evolution achieved by CNN-assisted DE, (e) CNN-assisted DE power evolution results for symmetric input plotted with its reversed version in linear scale (mW) (f) The asymmetry in frequency for the CNN-assisted DE.}
        \label{fig:symmetric results}
\end{figure*}

This 2D power profile satisfies the symmetric condition in distance. Additionally, at each distance point, it has the same power level for all frequencies. Table \ref{symmetric poump powers} shows the set of pump power values achieved by both the CNN model and the CNN-assisted DE technique. It should be noted that all hyper-parameters required for the optimization have been set as we did for 2D flat power evolution design. Fig. \ref{fig:symmetric results} shows the results for 2D symmetric power profile design. 

\begin{table}[ht]
\caption{Predicted pump power values by CNN and CNN-assisted DE optimization for symmetric power evolution profile}
\centering
\label{symmetric poump powers}
\begin{tabular}{|l||*{2}{c|}}\hline

    \hline
    \backslashbox{Pump}{model}
    &\makebox{CNN only}&\makebox{CNN-assisted DE}\\\hline\hline
    $p_1[mW]$ & 150 & 100 \\ [3pt]\hline
    $p_2[mW]$ & 6 & 7 \\[3pt] \hline
    $p_3[mW]$ & 63 & 36 \\[3pt] \hline
    $p_4[mW]$ & 13 & 7 \\ [3pt]\hline
    $p_5[mW]$ & 1230 & 1480 \\[3pt] \hline
    $p_6[mW]$ & 20 & 13 \\ [3pt]\hline
    $p_7[mW]$ & 10 & 6 \\ [3pt]\hline
    $p_8[mW]$ & 72 & 33 \\[3pt] \hline

\end{tabular}
\end{table}

Fig. \ref{fig:symmetric results} (a) illustrates the power evolution resulting from the pump power values predicted for symmetric input with the dashed black curve showing the symmetric target profile generated based on Eq. \ref{eq:sin_wave}. To have a visual understanding of the asymmetry minimization of power evolution defined in the linear scale [mW], we have shown the resulting 2D power evolution together with its reversed version in distance in Fig. \ref{fig:symmetric results} (b). Additionally, the asymmetry value has been shown as a function of frequency in Fig. \ref{fig:symmetric results} (c) asserting the maximum asymmetry of 33.2 \% at 195.9 THz, and the minimum asymmetry of 28.5 \% at 192.1 THz. 

Furthermore, Fig. \ref{fig:symmetric results} (d) shows the power profile resulted from CNN-assisted DE with the dashed black curve indicating the symmetric target power evolution for all channels. Since the optimization process in Eq. \ref{eq:sym1_opt} is performed to minimize asymmetry with power evolution defined in linear scale [mW] (Eq. \ref{eq:sym1}), the resulting power evolution and the reversed version, both in linear [mW] scale, are shown in Fig. \ref{fig:symmetric results} (e). It is visually obvious that the asymmetry has improved strongly by applying DE optimization compared to its reversed version. Eventually, Fig. \ref{fig:symmetric results} (f) shows the asymmetry improvement for all frequencies with maximum value of 14\% over the entire C-band, asserting more than 20\% decrease in average asymmetry value by applying the fine-tuning process. Additionally, based on Fig. \ref{fig:symmetric results} (f), the resulting maximum 14\%  asymmetry is identified at 195.7 THz, and the minimum asymmetry is 7.2 \%, which is identified at 193.8 THz.

\begin{figure}[H]
    \centering
    \includegraphics[width = 8.5cm]{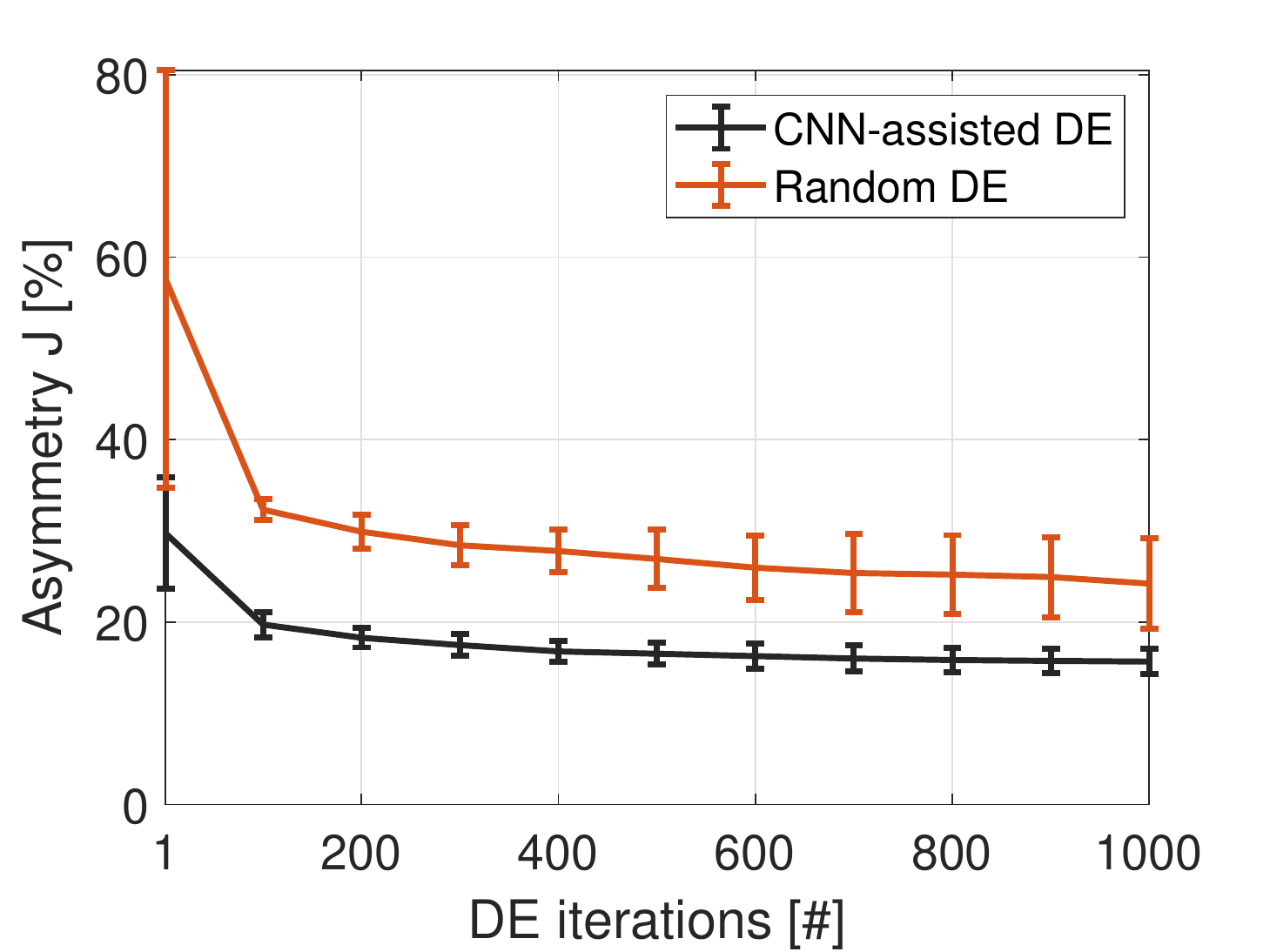}
    \caption{Average asymmetry (marker) and standard deviation (error bars) for the CNN-assisted DE and the DE with random initialization as a function of number of DE iterations.}
    \label{fig:symmetric error bar}
\end{figure}

Similar to the analysis performed for flatness, we consider two pump power adjustment approaches to illustrate the impact of the CNN model used to initialize the DE population. In the first approach, the DE is assisted by the CNN model and in the second one, the CNN is not involved and the DE fine-tuning is performed by setting the $\textbf{p}_{LB}$ and $\textbf{p}_{UB}$ to the minimum and maximum values of pump powers, respectively.

 Fig.\ref{fig:symmetric error bar} shows the line plot as the average error with the error bars for these cases indicating almost 10\% average cost improvement when CNN is involved in the population initialization. Moreover, it is obvious that the optimization without using CNN referred to as random DE has higher standard deviation at most of the points and this value increases with the number of evaluations. Hence, the probability of convergence to different minimum values is higher in case of running the DE algorithm blindly and without utilizing CNN to specify the initial solution.

\section{Conclusion}
We proposed a CNN-assisted DE framework to adjust Raman pump power values for designing 2D target power profiles over the spectral and spatial domains in the fiber span. In the proposed framework, a pre-trained CNN model predicts the set of pump power values for a 2D target profile, and a DE algorithm performs gradient-free fine-tuning process aiming to improve the predicted values by the CNN. The CNN is trained on the data generated by exciting the amplification setup with randomly selected pump powers. However, its performance is not accurate when the target is to design specific 2D profiles of practical interest. To improve the pump powers predicted by the CNN, the DE employs a direct Raman solver function and performs fine-tuning to predict a new set of pump powers resulting in a feasible 2D profile with lower cost. The results assert the very good performance of the proposed CNN-assisted DE framework utilized for 2D flat and symmetric power profiles design. The proposed framework achieves power excursion of 2.81 dB for a flat target, and maximum asymmetry of 14\% for a symmetric target in a 80 km fiber span in the whole C-band. Furthermore, the proposed DE with the CNN initialization provides higher accuracy with lower variance compared to the randomly initialized DE optimization.


%

\section*{Acknowledgment}
This work was financially supported by the European Research Council (ERC-CoG FRECOM grant no. 771878), the Villum Foundation (OPTIC-AI grant no. 29334), and the Italian Ministry for University and Research (PRIN 2017, project FIRST).

\ifCLASSOPTIONcaptionsoff
  \newpage
\fi



%

\bibliography{ref}
\bibliographystyle{ieeetr}

\end{document}